\documentclass[intlimits,twoside,a4paper]{article}
\usepackage{amsmath,amssymb}
\usepackage{graphicx}
\usepackage{wrapfig}
\usepackage[T2A]{fontenc}
\usepackage[cp1251]{inputenc}

\usepackage{cmpj3}

\issue{2020}{23}{1}{13604}
\doinumber{10.5488/CMP.23.13604}

\title[Dislocation loops growth]
{Dislocation loops growth and radiation growth in neutron irradiated Zr-Nb alloys: rate theory modelling}

\author[L. Wu \textsl{et al.}]{L. Wu\refaddr{label1}, D.O.~Kharchenko\refaddr{label2}, V.O.~Kharchenko\refaddr{label2}, O.B.~Lysenko\refaddr{label2}, V. Kupriienko\refaddr{label2}, S.~Kokhan\refaddr{label2}, I.A.~Shuda\refaddr{label3},  R. Pan\refaddr{label1}}
\addresses{
\addr{label1} The First Institute, Nuclear Power Institute of China, 328, the 1st
Section, Changshundadao Road, Shuangliu, 610213  Chengdu, China
\addr{label2} Institute of Applied Physics of the National Academy of Sciences of Ukraine,\\ 
58 Petropavlivska St., 40000 Sumy, Ukraine
\addr{label3} Sumy State University, 2 Rymskogo-Korsakova St., 40007
Sumy, Ukraine
} 

\date{Received May 29, 2019, in final form October 23, 2019}

\begin{document}
\maketitle
\begin{abstract}
A generalized model to study dislocation loops growth in irradiated  binary  Zr-based alloys  is presented. It  takes into account temperature effects, efficiencies of loops to absorb point defects dependent on the loop size, an influence of locality of grain boundary sink strength, and concentration of the alloying element. This model is used to describe the dynamics of loop radii growth in zirconium-niobium alloys under neutron irradiation at  reactor conditions. A growth of both loop radii and strains  is studied at  different grain sizes,  location from  grain boundaries, and concentration of niobium. It is shown that locality of grain boundary sinks results in a non-uniform deformation of the crystal inside the grains. Additionally, an introduction of niobium as an alloying element decreases the loop radii but promotes the growth of local strains inside the grains.

\keywords  irradiation, zirconium alloys,  point defects, dislocation loops, growth strain
\end{abstract}

\section{Introduction}

Zirconium alloys having low capture cross section to thermal
neutrons and good corrosion resistance are used as cladding materials in structural components for light and heavy water nuclear reactor cores.
It was shown experimentally that  zirconium-based alloys are characterized by  excellent mechanical properties,  superior corrosion resistance due
to the formation of a passive oxide on the surface~\cite{AKM12,MCC15,MKMSYMYTYWKMYLSKA17}.
At the same time, these  alloys manifest  dimensional instabilities in
a high-temperature irradiation environment of a nuclear reactor,
leading to irradiation growth, irradiation
enhanced creep and accumulation of hydrogen~\cite{G08,H08,McG11}.

Zirconium-based  alloys  are usually subjected
to a fast neutron flux with neutron energy $1{-}2$~MeV leading to 
irradiation damages  due to elastic interaction between fast neutrons and atoms
of the alloy. It leads to  an atoms  displacing from their positions, creating cascades of atomic displacements and generating point defects. 
During cascade annealing more than $90\%$ of defects relax onto sinks (loops, grain boundaries, phase interfaces, cavities) and a remaining part of defects is observed as isolated or clustered. Usually,  interstitial clusters contain up to 25 interstitials, vacancy clusters are of 24 vacancies at 600~K at the
end of the cascade relaxation \cite{Hodgson,GBHS01}.

\looseness=-1 Clusters of point defects  grow by absorbing  unrelaxed defects from the bulk transforming into dislocation loops of $\langle a\rangle$ and  $\langle c\rangle$ types \cite{koenig1}.
Usually $\langle a\rangle$-type loops (in prismatic planes with Burgers vector $\langle a\rangle=1/3\langle11\overline{2}0\rangle$) are either of vacancy or interstitial character. Their typical diameter is  of $5{-}20$~nm at doses up to several dpa (displacement per  atom).
As was shown in \cite{koenig1,NGB79,G88}, the proportion of vacancy loops to interstitial
loops depends on the irradiation temperature: at  temperature
of $633$~K one has  50\% of  vacancy loops; at  temperature
of $673$~K vacancy loops are of 70\%.
At temperatures below $573$~K, most of the loops are of the interstitial type.
A simultaneous observation of vacancy and
interstitial $\langle a\rangle$-type loops is reported in the literature \cite{NGB79,KB73,JKB77,GGF87,G93}.
As was pointed out in \cite{BGS15}, the co-existence of vacancy type and interstitial type loops violates the well-known loop property of the same efficiencies for absorption of
point defects.
 Dislocation $\langle c\rangle$-type loops (in basal planes with Burgers vector $1/6\langle 20\overline{2}3\rangle$) 
 are of vacancy type only, they  start to form at doses up to 3~dpa. Their diameter is larger than the diameter of  $\langle a\rangle$ loops, but the density is much lower \cite{koenig1}.
Transmission electron
microscopy (TEM) studies indicate that $\langle a\rangle$ loop number
density saturates during irradiation attaining the typical value $10^{22}$~m$^{-3}$~\cite{CN75}.
Dislocation $\langle c\rangle$-type loops  attain a stationary value of the loop  number density $10^{21}$~m$^{-3}$ \cite{HG86,GG87}. 
Large faulty vacancy
dislocation $\langle c\rangle$ loops  and $\langle a\rangle$ loops are responsible for the accelerated process of irradiation growth as a process with 
an emergence of internal strains in the crystal when growth strains in all crystallographic directions are nonzero but the total volume of the system remains constant \cite{RRT,GGF89,RM83}. 
Experimental studies show that the irradiation growth  depends  on neutron fluence \cite{HG86,Zinkle}, temperature
\cite{GGF87,GGF89,Zinkle}, alloy composition \cite{TFA84} and thermomechanical history  \cite{AT82,R88,Zinkle} but a detailed mechanistic understanding is
presently absent.

Models for irradiation growth with 
different mechanisms were proposed in the literature  to describe the changes of material properties  during irradiation of single- and poly-crystals \cite{BW80,SL80,MC80,H80,H88,FMSS94,W98}.
Theoretical description of the radiation growth was developed by taking into account a diffusion anisotropy difference effect where diffusion of self-interstitials  is anisotropic, whereas vacancy diffusion can be considered isotropic~\cite{F88,B88,OBD02,WHZ03,WL07}. Cluster dynamics approach proposed in \cite{HMSB02,CB05} to study the radiation growth was exploited to describe experimental data and predict a  change of mechanical properties of zirconium single crystals  and Zircaloy-2 under neutron irradiation \cite{CB09,GBS13,BGS15,PTG17}. In  models of point  defect clustering one considers the kinetics of defect
production with their interactions with the dislocation substructure. It allows one to predict the dynamics of growth strains. In \cite{GBS13,BGS15}, a simplified approach was introduced to study the dynamics of dislocation loops. It was shown that loops start to evolve  from defect clusters where a net flux of defects is considered as an additional mechanism for the growth of the loops. In \cite{PTG17}, the  irradiation growth was studied in the framework of viscoplastic self-consistent crystal plasticity in Zr and its alloys. This approach is based on a response of individual grains interacting with the surrounding medium. Here, an anisotropy,  grain interactions, grain size and grain shape effects are taken into account to predict the radiation growth. 

In theoretical description of dislocation loop dynamics there exist open problems related to an influence of the efficiencies of dislocation loops of different type  to absorb  interstitials and vacancies. On the other hand, by considering the binary alloys, one should take into  account that  precipitates with their phase interfaces  ($\beta$-phase particles of niobium in Zr-$x$\%Nb alloys) serve as additional sinks of point defects: the phase interface between $\alpha$-zirconium and $\beta$-niobium accumulates the point defects acting as a grain boundary. 
Moreover, 
one should take into account the locality of a sink strength related to grain boundaries when  a diffusion of point defect clusters is considered. As was pointed out in \cite{RRT,SGTSOB97}, the
sink strength of grain boundaries  depends on the distance of the local area to the grain boundary. 

\enlargethispage{2pt}

In this work, we focus on the problem of dislocation loop dynamics in a  model of point defect and evolution of their clusters by taking into account the grain boundary and precipitates as additional sinks for point defects with   temperature effects in the framework of reaction rate theory \cite{RRT}. The last one can be achieved when recombination processes of point defects  are considered and a steady state balance condition is satisfied. In our model we take into account the bias of dislocation loops to absorb the defects and bias factors for grain boundaries and $\beta$-phase precipitates. As was shown previously (see, for example,~\cite{RRT}), the efficiency  of dislocation loops to absorb point defects of different type decreases with the loop radius growth. We take into account this effect to describe the growth of the loops more accurately at the stage of the growth at small sizes where one assumes that loops exist separately and do not interact. At large loop sizes, these efficiencies are assumed to be constant. A novelty of this work is in (i) accounting bias factors for sinks of point defects and their clusters and temperature effects,  (ii) local description of the grain boundary influence onto absorption  of point defect clusters, leading to local description of both dislocation radii evolution and growth strains development inside the grain, (iii) possibility to describe the growth of the loops and strains   by varying concentration of niobium as an alloying element.

The paper is organized in the following manner. In section~\ref{sec2} we describe a general model for binary alloys where two types of interstitials are possible. Here, the sinks of point defects are: network dislocations, loops, grain boundaries and precipitates of the secondary phase. We assume that  both interstitial and vacancy clusters emerge during the development of a defect structure. We exploit this model to describe the dislocation loops growth in  Zr-$x$\%Nb alloys basing on a set of material parameters and assumptions used in works \cite{GBS13,BGS15,PTG17} for the dynamics of dislocation loop number densities. The results of numerical modelling within  this approach are presented in section~\ref{sec3}. Here, we study an influence of sizes of grains, precipitates, content of niobium and temperature at a fixed dose rate.  We conclude in  section~\ref{sec4}.

\section{The model and rate equations} \label{sec2}
\subsection{General description}
Let us consider a binary alloy where the concentrations ${c_\text{A}}$ and ${c_\text{B}}$ of atoms of both sorts A and B obey the conservation law ${c_\text{A}}+{c_\text{B}}=1$. 
It is assumed that this alloy is subjected to sustained irradiation with the dose rate  computed according to  the Norgett-Robinson-Torrens (NRT) standard \cite{NRT}: $\mathcal{K}_\text{NRT}\approx\sigma_\text{d}\nu_\text{FP}\Phi$, where $\sigma_\text{d}$ is  the neutron cross-section, $\nu_\text{FP}$ is the number of Frenkel pairs  generated in cascades, $\Phi$ is  the total particle flux.  By taking into account a fraction of defects recombined in cascades, $\epsilon_\text{r}$, next it is convenient to use the shortest notation for the damage rate, $\mathcal{K}=\mathcal{K}_\text{NRT}(1-\epsilon_\text{r})$. It is known that interstitials and vacancies  produced in cascades can be clustered with efficiencies $\varepsilon_\text{i}$ and $\varepsilon_\text{v}$, respectively; some of them relax on sinks representing free surfaces,  grain boundaries or interfaces, network dislocations and dislocation loops. The last kind of sinks is  formed in the processes of the growth of  point defect clusters.   The total amount of interstitials can be expressed as $c_\text{i}=c_\text{i}^\text{A}+c_\text{i}^\text{B}$, where $c_\text{i}^\text{A,B}$ denotes concentration of interstitials of each sort. We denote the vacancy concentration by  $c_\text{v}$.

To describe the dynamics of the defects produced in cascades at large time scales,  we exploit the main ideas of the reaction rate theory \cite{RRT}: interstitials of each sort are produced from  atoms of the corresponding sort, vacancies  are generated in cascades when atoms leave their positions. Rates of these processes are defined by the damage rate $\mathcal{K}$. Annihilation processes are described by the rates defined through  point defect sinks and  diffusion coefficients.  Recombination processes are taken into account for each sort of interstitials.

If some macroscopic heterogeneities (nonhomogeneous distributions of the sinks, presence of external forces or influence of the boundary) exist, then  the macroscopic diffusion processes occur in the system that may change the spatial defect distributions. Therefore, in general case one arrives at reaction-diffusion kinetics by introducing corresponding diffusion fluxes. 
It allows one to write down dynamical equations for point defects as follows:
\begin{equation}
\begin{split}
\partial_t  &c^\text{A,B}_\text{i}={\mathcal{K}}(1-\varepsilon_\text{i}){c_\text{A,B}}-D^\text{A,B}_\text{i}(k^2_\text{i}+k^2_\text{GB}+k^2_\text{p}) c_\text{i}^\text{A,B}-\alpha^\text{A,B} c^\text{A,B}_\text{i}c_\text{v}-\nabla\cdot \mathbf{J}_\text{i}^\text{A,B}.
\end{split}
\end{equation}
Here, $D_\text{i}^\text{A,B}$  denotes  diffusivities of each sort of interstitials,   $k^2_{\{\cdot\}}$ denotes a sink strength of network dislocations, dislocation loops, grain boundaries and precipitates. We define the sinks strength for interstitials as 
$k^2_\text{i}=\sum_jk^2_{\text{i};j}$, where the sum is provided over all crystallographic directions $j$,  $k^2_{\text{i};j}\equiv Z_{\text{iN}}^j\rho_\text{N}^j+Z_{\text{iV}}^j\rho^j_\text{V}+Z_{\text{iI}}^j\rho^j_\text{I}$ takes into account 
network dislocation density $\rho_\text{N}^j$,  interstitial and vacancy (I/V) loops with densities $\rho^j_\text{I,V}=2\piup R_\text{I,V}^jN^j_\text{V}$,  where $R_\text{I,V}^j$  and $N^j_\text{I,V}$ denote the corresponding loop radius and loop number density, respectively; $k^2_\text{GB}=\sum_j\sum_nZ^{j}_{n\,\text{GB}}(\lambda_{n\,\text{GB}})^{-2}$ relates to grain boundaries  of the  radius $\lambda_{n\,\text{GB}}$ in one of three Cartesian directions $n=(\mathbf{x}, \mathbf{y},\mathbf{z})$; $k^2_\text{p}=\sum_j\sum_nZ^{j}_{n\,\text{p}} 4\piup r_\text{p}N_\text{p}$ is the sink strength of precipitates of the size $r_\text{p}$ and the density $N_\text{p}$;
 $Z_{\{\cdot\}}^j$ denotes  bias factors which originate from the difference in the point defect motion to dislocations, loops, grain boundaries and phase interfaces;  $\alpha^{\{\cdot\}}$ relate to   the corresponding recombination rate constants.
 In our description we take into account the standard definitions
$\alpha^\text{A,B}=2\piup{(D^\text{A,B}_\text{i}+D_\text{v}^\text{A,B})r^\text{A,B}_\text{c}}/{\Omega^\text{A,B}}$, where $r^\text{A,B}_\text{c} \simeq(2-3)a^\text{A,B}$  relates to capture radius, $a^\text{A,B}$ is the effective lattice parameter defined by the alloy composition, $D_\text{v}^\text{A,B}$ relates to vacancy diffusivity in each phase (A or B), $\Omega^\text{A,B}$ is the corresponding atomic volume. If a lattice mismatch is small enough, then one can put  $a^\text{A}\simeq a^\text{B}=a$ and assume   $r_\text{c}^\text{A}\simeq r_\text{c}^\text{B}=r_\text{c}$, $\Omega^\text{A}\simeq \Omega^\text{B}=\Omega$.
The diffusion fluxes are introduced in the standard form as follows: $\mathbf{J}^\text{A,B}_\text{i}=-D_\text{i}^\text{A,B}\nabla c_\text{i}^\text{A,B}$.

Let us assume that A-component is the host, whereas B represents an alloying component. In the mean field approximation, we consider the averaged concentration of atoms and take into account that the alloying component has the averaged concentration $x\equiv \overline{c_\text{B}}$, whereas $\overline{c_\text{A}}=1-x$. Therefore, for interstitials  one can use the following relations obeying the above laws for the sum of point defects: $c_{\text{i}}^\text{A}=(1-x)c_{\text{i}}$, $c_{\text{i}}^\text{B}=xc_{\text{i}}^\text{B}$, $c_\text{i}=c_\text{i}^\text{A}+c_\text{i}^\text{B}$. Next, by combining two equations for interstitials,  one gets the reduced system describing the dynamics of the total amount of interstitials and vacancies:
\begin{equation}\label{pdef}
\begin{split}
\partial_t  &c_{\text{i}}={\mathcal{K}}(1-\varepsilon_\text{i})- D_\text{i}(x)(k_\text{i}^2+k^2_\text{GB}+k^2_\text{p})c_{\text{i}}-\alpha(x)c_{\text{i}}c_\text{v}-\nabla\cdot\mathbf{J}_\text{i}\,,\\
\partial_t  &c_\text{v}={\mathcal{K}}(1-\varepsilon_\text{v})- D_\text{v}(x)(k^2_\text{v}+k^2_\text{GB}+k^2_\text{p})(c_\text{v}-c_{0\text{v}})-\alpha(x) c_{\text{i}}c_\text{v}-\nabla\cdot\mathbf{J}_\text{v}.
\end{split}
\end{equation}
Here, $k^2_\text{v}=\sum_jk^2_{\text{v};j}$ is the sink strength of dislocations for vacancies, where  $k^2_{\text{v};j}\equiv Z_\text{vN}^j\rho_\text{N}^j+Z_\text{vV}^j\rho^j_\text{V}+Z_\text{vI}^j\rho^j_\text{I}$. 
The concentration dependent  diffusivities  and recombination rate are: 
\begin{equation}
\begin{split}
D_\text{i}(x)&=D^\text{A}_{\text{i}}(1-x)+D^\text{B}_{\text{i}}x;\\D_\text{v}(x)&=D^\text{A}_\text{v}(1-x)+D^\text{B}_\text{v}x;\\ 
\alpha(x)&=\alpha^\text{A}(1-x)+\alpha^\text{B}x.
\end{split}
\end{equation}
Last terms in equation~(\ref{pdef}) relate to the corresponding diffusion fluxes for interstitials and vacancies, respectively. It is known that diffusion processes play an essential role in large systems or at localization of sinks (their inhomogeneous spatial distribution), when a  mean free path of diffusion species is of the order of the distance  between their sinks. For point defects, as usual, this value is around the distance between network dislocations (or a grain size),   $l_\text{D}\propto (\rho_\text{N}+k^2_\text{GB})^{-1/2}\approx 1{-}10$~{\textmu}m. 
In the system under consideration one takes into account the dislocation loops and precipitates homogeneously distributed over the whole system. It results in a decrease in the mean free path of point defects which can be captured by sinks on the distances less than the diffusion length. To estimate the change of the mean free path, one can consider the typical values for dislocation density $\approx10^{12}$~m$^{-2}$, grain size $\approx 10^{-5}$~m, dislocation loop density $\approx  10^{14}$~m$^{-2}$ and size of $\beta$-phase Nb precipitate $\approx 5\times 10^{-9}$~m with their number density $10^{19}{-}10^{20}$~m$^{-3}$. It follows that the mean free path distance of the point defects with all the above sinks is $l_\text{s}\propto(k_\text{i}^2+k_\text{v}^2+k^2_\text{GB}+k^2_\text{p})^{-1/2}\approx 0.1$~{\textmu}m which is less (at least by an order) than distance between dislocations defining the diffusion length of point defects, $l_\text{s}\ll l_\text{D}$. It means that point defects will be absorbed by sinks from the bulk  even when they start to diffuse.  In this case, the long-range diffusion of point defects becomes negligibly small compared to the influence of the sinks. It allows us to use zero-order approximation and consider homogeneous defects distribution by neglecting the corresponding diffusion fluxes \cite{Was}.

By considering the dynamics of interstitial and vacancy clusters, we follow the works \cite{RRT,SGTSOB97} where it was shown that the efficiency of absorbing clusters by grain boundaries depends on the distance between clusters and  grain boundaries as their main sinks. In the case under consideration, the dynamics of  cluster concentrations $c^{\,j}_{\text{i},\text{v}\,\text{cl}}$ produced in $j$-th crystallographic direction is described by the following equations:
\begin{equation}\label{clust}
\begin{split}
\partial_t  &c^{\,j}_{\text{i}\,\text{cl}}=\frac{\mathcal{K}\varepsilon_\text{i}}{d_\text{i}n_\text{i}}-D^j_{\text{i}\,\text{cl}}S^j_{\text{i}\,\text{cl}}(\ell)c^{\,j}_{\text{i}\,\text{cl}}\,,\\
 \partial_t &c^{\,j}_\text{v\,cl}=\frac{\mathcal{K}\varepsilon_\text{v}}{d_\text{v}n_\text{v}}-D^j_\text{v\,cl}S^j_\text{v\,cl}(\ell)c^{\,j}_\text{v\,cl}.
\end{split}
\end{equation}
Here, $d_{\text{i},\text{v}}$ denotes the number of directions along which clusters can be formed;  $n_{\text{i},\text{v}}$ is the number of interstitial or vacancies in the corresponding cascade-produced mobile cluster; $D_{\text{i,v}\,\text{cl}}$ relates to diffusivity of clusters; $S^{j}_{\text{i},\text{v\,cl}}$ is the sink strength  depending on the distance of a local area to the grain boundary, $\ell$. For the  sink strength of grain boundaries,  we  use a definition proposed in works \cite{RRT,SGTSOB97}:
$$S^{j}_{\text{i},\text{v\,cl}}(\ell)=2\left( \frac{\piup r_{\text{i},\text{v\,cl}}\tilde \rho^j_{\text{i},\text{v}}}{2}\right)^2,\qquad 
\tilde\rho^j_{\text{i},\text{v}}=\rho^{\,j}+\frac{2}{\piup r_{\text{i},\text{v\,cl}}}\frac{1}{\sqrt{\ell(2\lambda_\text{GB}-\ell)}}\,,$$ 
where  $r_{\text{i},\text{v\,cl}}$ is the capture radius of  loops for mobile clusters, $\rho^j=\rho_\text{N}^j+\rho^j_\text{V}+\rho^j_\text{I}$.
 Here and thereafter, it is assumed that $\lambda_\text{n\,GB}=\lambda_\text{GB}$, for simplicity.

Let $x^j_\text{I,V}=\piup (R^{j}_\text{I,V})^2b^{\,j} N^j_\text{I,V}$ be the total number of defects in interstitial/vacancy (I/V)-loops of the size $R^j_\text{I,V}$, $b^{\,j}$ be the modulus of the Burgers vector.
An evolution of the total number of defects  in the loops is described by the following equations:
\begin{equation}
\begin{split}\label{x_iv}
\partial_t x^j_\text{I}&=-2\piup R_\text{I}^j N_\text{I}^j \big[Z_\text{vI}^jD_\text{v}(x) (c_\text{v}-c_{\text{iL}})-Z_{\text{iI}}^jD_\text{i}(x)c_\text{i}\big]+S^j_\text{I}(n_\text{i}D^j_\text{cl}c^j_{\text{i\,cl}}-n_\text{v}D^j_\text{v\,cl}c^j_\text{v\,cl});\\
\partial_t x^j_\text{V}&=2\piup R_\text{V}^j N_\text{V}^j \big[Z_\text{vV}^jD_\text{v}(x) (c_\text{v}-c_\text{vL})-Z_{\text{iV}}^jD_\text{i}(x)c_\text{i}\big]-S^j_\text{V} (n_\text{i}D^j_{\text{i\,cl}}c^j_{\text{i\,cl}}- n_\text{v}D^j_\text{v\,cl}c^j_\text{v\,cl});\\
\end{split}
\end{equation}
where, following the receipt discussed in \cite{BGS15,PTG17}, one has
$$S^j_\text{I,V}=(\piup r_{\text{i},\text{v\,cl}})^2(\piup R_\text{I,V}^j N_\text{I,V}^j)\tilde\rho^{\,j}_{\text{i,v}};$$
the concentrations of vacancies which are in equilibrium with vacancy- and interstitial-type loops are: 
\begin{equation}\label{c_{ivL}}
c^j_\text{vL}=c_\text{v0}\exp\left[\frac{(\gamma_\text{SF}^j+F^\text{e}_{\text{V};j})(a^{j})^2}{T}\right],\qquad c^j_{\text{iL}}=c_\text{v0}\exp\left[-\frac{(\gamma_\text{SF}^j+F^\text{e}_{\text{I};j})(a^{j})^2}{T}\right];
\end{equation}
$\gamma_\text{SF}^j$ is the stacking fault energy, the quantity $F^\text{e}_{\text{V,I};j}=\frac{\mu (b^{j})^2\ln(R^j_\text{V,I}/a^j+1)}{4\piup(1-\nu)(R^j_\text{V,I}+a^j)}$ relates to elastic energy stored in loops; $\mu$ is the shear modulus, $\nu$ is the Poisson ratio, as usual, $a^j$ is the lattice constant in  direction $j$.

By using equations~(\ref{x_iv}) one can get the evolution equations for the loop  radii $R^j_\text{I,V}$ or for the loop densities $\rho^j_\text{I,V}$ (the last one is possible under assumption that $N^j_\text{I,V}=\text{const}$ valid at large doses). In general, the loop number densities $N^j_\text{I,V}$ depend on time, therefore, in order  to make a closed loop system, one needs to introduce a model for the dynamics of $N^j_\text{I,V}$ by using the data of experimental observations of the formation of the loops.

\subsection{Defect dynamics in Zr-Nb alloy}

In Zr-Nb alloys studied below one takes into account  that most of Nb (from 82\% up to 90\% of its content) is in $\beta$ (bcc) phase of the size around 5--40~nm,  the rest is in the solid solution characterized by $\alpha$ phase of hcp structure; $\
\beta$-phase precipitates are localized mostly at grain boundaries of $\alpha$-zirconium matrix \cite{35}.  Therefore, the above  mean field approach dealing only with the averaged values of point defect concentrations, dislocation loop radii, and alloy constituent concentrations is useful mostly for solid solution with small niobium content inside the grains of zirconium matrix  characterized by four crystallographic directions,  $j=(\mathbf{a}1, \mathbf{a}2, \mathbf{a}3, \mathbf{c})$.  In such a case, dislocation loops will be  characterized by the Burgers vector lying in prismatic directions  $\mathbf{a}1$, $\mathbf{a}2$, and, $\mathbf{a}3$ and in the basal plane ($\mathbf{c}$). It is taken into account that  basal interstitial-type loops are not formed.

At the same time, an influence of phase interfaces of $\beta$-Nb precipitates should be taken into account by using the concentration of niobium  in the alloy and its content in 
$\beta$-precipitates. To set the appropriate definition of the sink strength $k_\text{p}^2$ depending on $x$, one can compute the maximal number density  $N_\text{p}^\text{max}=\Delta P_\text{p}/V_\text{GB}$ of  precipitates of the size $r_\text{p}$ located at the grain boundary of the radius $\lambda_\text{GB}$ (along $\alpha{-}\alpha$~interfaces), where  $\Delta P_\text{p}=P^\text{tot}_\text{p}-P^\text{in}_\text{p}$ is the number of precipitates, $V_\text{GB}$ is the  volume of the spherical grain.
The total number of precipitates closely packed in the spherical grain is $P^\text{tot}_\text{p}=\frac{\piup}{3\sqrt{2}}(\lambda_\text{GB}/r_\text{p})^3$, where $\frac{\piup}{3\sqrt{2}}$ is the Gaussian number for close packing of spheres. The number of  precipitates closely packed   in a sphere of the size $\lambda_\text{GB}-r_\text{p}$ is $P^\text{in}_\text{p}=\frac{\piup}{3\sqrt{2}}[(\lambda_\text{GB}-r_\text{p})/r_\text{p}]^3$. 
By taking into account that $\lambda_\text{GB}/r_\text{p}\gg 1$, one finds $\Delta P_\text{p}=\frac{\piup}{3\sqrt{2}}[1+3(\lambda_\text{GB}/r_\text{p})(\lambda_\text{GB}/r_\text{p}-1)]\approx\frac{\piup}{\sqrt{2}}(\lambda/r_\text{p})^2$.
Therefore, the maximal number density of Nb-precipitates located at the grain boundary  is $N_\text{p}^\text{max}\approx\frac{3}{4\sqrt{2}}(\lambda_\text{GB}r_\text{p}^2)^{-1}$. As far as the content of  niobium in $\beta$-precipitates  is about 0.9, one can compute  the real number of precipitates  $N_\text{p}=N_\text{p}^\text{max}\times 0.9x$.
An estimation of $N_\text{p}$ at $r_\text{p}=5$~nm, $\lambda_\text{GB}=10$~{\textmu}m, $x=0.025$ gives $N_\text{p}^\text{max}=2.12\cdot 10^{15}$~cm$^{-3}$ and $N_\text{p}=4.8\cdot10^{13}$. Experimental observations (see \cite{XXX} and citations therein) for the number densities of $\beta$-Nb particles give  $N_\text{p}\sim 10^{13}{-}10^{14}$~cm$^{-3}$. Therefore, the provided estimations relate well to experimental data. Small corrections can be done by including temperature dependencies of $r_\text{p}$ and $N_\text{p}$.

Point defects migrate in 3D lattice. Interstitial clusters produced in cascades  are characterized by the Burgers vector along prismatic directions \cite{koenig1,BGS15}. According to experimental observations, it follows that  interstitial clusters execute 1D migration along the Burgers vectors parallel to the basal planes. We assume that vacancy clusters are formed in all planes. They are  mobile in all directions. It is taken into account that point defect clusters interact with dislocations of the same Burgers vector only \cite{BGS15}.

The dynamics of the system of defects will be described  by concentration of point defects according to  equation~(\ref{pdef}). The dynamics of interstitial clusters in each  prismatic direction $m=(\mathbf{a}1, \mathbf{a}2, \mathbf{a}3)$ are equivalent. For vacancy clusters, we assume that all four directions are equivalent. Hence, to describe clusters of point defects  one can put $D_{\text{i\,cl}}\equiv D^m_{\text{i\,cl}}$, $D_\text{v\,cl}\equiv D^j_\text{v\,cl}$, $S^m_{\text{i\,cl}}\equiv S_{\text{i\,cl}}$, $S_\text{v\,cl}\equiv S^j_\text{v\,cl}$,  and $d_\text{i}=d_\text{v}=3$ in equation~(\ref{clust}). In our consideration, we have an  interest in a long-time  behaviour. By neglecting initial transients in the concentration of point defects and their clusters,  a steady state balance can be assumed, $\partial_t c_\text{v}\approx\partial_t c_\text{i}\approx\partial_t c_{\text{i\,cl}}\approx\partial_tc_\text{v\,cl}\approx 0$. 

By taking into account that interstitial loops are possible in prismatic directions only, an evolution equation for loop radius in one of prismatic directions can be obtained directly from equation~(\ref{x_iv}):
\begin{equation}
b_m\partial_t R^m_\text{I}=-[Z_\text{vI}^mD_\text{v}(x) (c_\text{v}-c^m_{\text{iL}})-Z_{\text{iI}}^mD_\text{i}(x)c_\text{i}]+\frac{\mathcal{K}(\piup r_{\text{i\,cl}})^2\tilde\rho^m}{6}\left[\frac{\varepsilon_\text{i}}{S^m_{\text{i\,cl}}(\ell)}-\frac{\varepsilon_\text{v}}{S^m_\text{v\,cl}(\ell)}\right]-\frac{b^mR^m_\text{I}}{2N_\text{I}^m}\partial_t N_\text{I}^m.
\end{equation}
Vacancy loops  in prismatic and basal planes are described by equations of the form:
\begin{equation}
\begin{split}
b_m\partial_t R^m_\text{V}&= Z_\text{vV}^mD_\text{v}(x) (c_\text{v}-c^m_\text{vL})-Z_{\text{iV}}^mD_\text{i}(x)c_\text{i}-\frac{\mathcal{K}(\piup r_\text{v\,cl})^2\tilde\rho^m}{6}\left[\frac{\varepsilon_\text{i}}{S^m_{\text{i\,cl}}(\ell)}-\frac{\varepsilon_\text{v}}{S^m_\text{v\,cl}(\ell)}\right]-\frac{b^mR^m_\text{V}}{2N_\text{V}^m}\partial_t N_\text{V}^m,\\
b_c\partial_t R^c_\text{V}&=Z_\text{vV}^cD_\text{v}(x) (c_\text{v}-c^c_\text{vL})-Z_{\text{iV}}^cD_\text{i}(x)c_\text{i}+\frac{\mathcal{K}(\piup r_\text{v\,cl})^2\tilde\rho^c\varepsilon_\text{v}}{2S^c_\text{v\,cl}(\ell)}-\frac{b_cR^c_\text{V}}{2N^c_\text{V}}\partial_t N_\text{V}^c.
\end{split}
\end{equation}

To get a closed loop system, one needs to know an equation for the loop number density $N_\text{I,V}$ modelling the loop nucleation processes. To this end, one can use the experimental observation for nucleation of loops. It is known that  $\langle a\rangle$-loops nucleate from the beginning of irradiation and reach the typical values $N^{m,\text{max}}_\text{I,V}\sim 10^{16}$~cm$^{-3}$ after several dpa, whereas vacancy $\langle c\rangle$-loops nucleate above a  critical dose  $\phi_\text{v}^{c,0}\simeq3$~dpa and later attain the density  $N_\text{V}^{c,\text{max}}\sim 10^{15}$~cm$^{-3}$ \cite{24,25,MC80,27}.  The simplest model for evolution of the loop number density was proposed in \cite{BGS15}. Its modification by  fitting experimental data  from SIRIUS reactor was done in \cite{PTG17}.  A comparison made by authors of these works has shown that the simplest model adequately explain experimental data for the nucleation of the loops. Therefore, in our consideration, we use this approach, where:
\begin{equation}\label{Na}
\partial_t N^m_\text{I,V}= \left\{
\begin{array}{ll}
\displaystyle\frac{\mathcal{K}N^{m,\text{max}}_\text{I,V}}{3\phi^a_\text{max}(1-\epsilon_\text{r})}\,, \quad & N^m_\text{I,V}\leqslant N^{m,\text{max}}_\text{I,V}/3,\\
0,\quad\quad &\text{otherwise};
\end{array} \right.
\end{equation}
\begin{equation}\label{Nc}
\partial_t N^c_\text{V}= \left\{
\begin{array}{ll}
0, \quad &\phi< \phi^{c}_{0}\,,\\
\displaystyle\frac{\mathcal{K}N_\text{V}^{c,\text{max}}}{1-\epsilon_\text{r}}\frac{A\exp[A(\phi-\phi_0^c)/(\phi^c_\text{max}-\phi_0^c)]}{(\phi^c_\text{max}-\phi_0^c)[\exp(A)-1]}\,,\quad &\phi\geqslant\phi_0^c,~ N_\text{V}^c<N_\text{V}^{c,\text{max}},\\
0, \quad\quad &\phi\geqslant\phi_0^c,~ N_\text{V}^c\geqslant N_\text{V}^{c,\text{max}}.
\end{array}  \right.
\end{equation}
We measure the accumulated dose as $\phi=\mathcal{K}t/(1-\epsilon_\text{r})$.

To compute the strains in all directions, we calculate the climb velocity as a function of the net flux of defects arriving at  dislocations. We take into account that both line dislocations and dislocation loops climb by
absorbing the point defects. The climb velocity is given by \cite{BGS15,PTG17}:
\begin{equation}\label{vclimb}
\begin{split}
v^m_\text{climb}&=\frac{n_\text{i}D^m_{\text{i\,cl}}S^m_{\text{i\,cl}}(\ell)c^m_{\text{i\,cl}}-n_\text{v}D^m_\text{v\,cl}S^m_\text{v\,cl}(\ell)c^m_\text{v\,cl}}{b^m\rho^m}
-\rho_\text{N}^m\frac{Z_\text{vN}^mD_\text{v}(x)(c_\text{v}-c_\text{0v})-Z_{\text{iN}}^mD_\text{i}(x)c_\text{i}}{b^m\rho^m}\\
 &-\rho_\text{I}^m\frac{Z_\text{vI}^mD_\text{v}(x)(c_\text{v}-c_{\text{iL}})-Z_{\text{iI}}^mD_\text{i}(x)c_\text{i}}{b^m\rho^m}-\rho_\text{V}^m\frac{Z_\text{vV}^mD_\text{v}(x)(c_\text{v}-c_\text{vL})-Z_{\text{iV}}^mD_\text{i}(x)c_\text{i}}{b^m\rho^m};\\
v^c_\text{climb}&=-\frac{n_\text{v}D^c_\text{v\,cl}S^c_\text{v\,cl}(\ell)c^c_\text{v\,cl}}{b^c\rho^c}-\rho_\text{N}^c\frac{Z_\text{vN}^cD_\text{v}(x)(c_\text{v}-c_\text{0v})-Z_{\text{iN}}^cD_\text{i}(x)c_\text{i}}{b^c\rho^c}\\&-\rho_\text{V}^c\frac{Z_\text{vV}^cD_\text{v}(x)(c_\text{v}-c_\text{vL})-Z_{\text{iV}}^cD_\text{i}(x)c_\text{i}}{b^c\rho^c}.\\
\end{split}
\end{equation}
Here, we use notations: $\rho^m=\rho_\text{N}^m+\rho_\text{I}^m+\rho_\text{V}^m$, 
$\rho^c=\rho_\text{N}^c+\rho_\text{V}^c$. The first terms in equation~(\ref{vclimb}) are responsible for contribution of clusters whereas last ones relate to collision of dislocations and loops with point defects. The effects of nucleation of the loops are not considered due to their small contribution compared to the other terms as was shown in \cite{BGS15}. 

The strain rate can be obtained from the  Orowan equation
\begin{equation}\label{eclimb}
\partial_te^{\,j}_\text{climb}=\rho^{\,j}b^{\,j}v^j_\text{climb}.
\end{equation}
The contribution from grain boundaries and phase interfaces considered as an extension of grain boundaries  to the strain rate relates to the net flux of point defects to grain boundaries as follows \cite{PTG17}:
\begin{equation}\label{e_g}
\partial_te^{\,j}_\text{GB}=-\sum_nZ_{n\,\text{GB}}^{j}\lambda_\text{GB}^{-2}[D_\text{v}(x)(c_\text{v}-c_\text{0v})-D_\text{i}(x)c_\text{i}].
\end{equation}

The total strain rate is 
$
\partial_te^{\,j}=\partial_te^{\,j}_\text{climb}+\partial_te^{\,j}_\text{GB}
$. The so-called ``macroscopic'' strain rate is of the form
\begin{equation}
\partial_te^{nl}=\sum_{j}b_n^jb_l^j\partial_te^{j}, \quad n,l=\{x,y,z\},
\end{equation}
where $b_n^j$ and $b_l^j$ are components of the Burgers vector $\mathbf{b}^{\,j}$ in $\mathbf{n}$ and $\mathbf{l}$ directions.

Finally, the dynamics of growth strains is described by the system of two equations:  
\begin{equation}
\begin{split}
\partial_{t} e^m&=\frac{\mathcal{K}(\varepsilon_\text{i}-\varepsilon_\text{v})}{3}-\sum_nZ_{n\,\text{GB}}^{m}\lambda_\text{GB}^{-2}\left[D_\text{v}(x)(c_\text{v}-c_\text{0v})-D_\text{i}(x)c_\text{i}\right]\\
&-\rho^m_\text{N}\left[Z_\text{vN}^mD_\text{v}(x)(c_\text{v}-c_\text{0v})-Z_{\text{iN}}^mD_\text{i}(x)c_\text{i}\right]
-\rho_\text{I}^m\left[Z_\text{vI}^mD_\text{v}(x)(c_\text{v}-c_{\text{iL}})-Z_{\text{iI}}^mD_\text{i}(x)c_\text{i}\right]\\&-\rho_\text{V}^m\left[Z_\text{vV}^mD_\text{v}(x)(c_\text{v}-c_\text{vL})-Z_{\text{iV}}^mD_\text{i}(x)c_\text{i}\right];\\
\partial_{t} e^c&=-\mathcal{K}\varepsilon_\text{v}-\sum_nZ_{n\,\text{GB}}^{c}\lambda_\text{GB}^{-2}\left[D_\text{v}(x)(c_\text{v}-c_\text{v0})-D_\text{i}(x)c_\text{i}\right]\\
 &-\rho_\text{N}^c\left[Z_\text{vN}^cD_\text{v}(x)(c_\text{v}-c_\text{0v})-Z_{\text{iN}}^cD_\text{i}(x)c_\text{i}\right]
-\rho_\text{V}^c\left[Z_\text{vV}^cD_\text{v}(x)(c_\text{v}-c_\text{vL})-Z_{\text{iV}}^cD_\text{i}(x)c_\text{i}\right].
\end{split}
\end{equation}

The derived formalism  for loop radii dynamics and growth strains calculations  generalizes the approaches discussed in works \cite{BGS15,PTG17,FPSP78} to calculate the radiation growth in zirconium based alloys by taking into account the bias coefficients for each kind of loops, grain boundaries and distance from the grain boundaries.

If Cartesian axes $x$, $y$ and $z$ are 
along $a1$, $a3-a2$ and $c$-directions, respectively, then  the strains along the
principal axes are correctly described by \cite{BGS15,PTG17}
\begin{equation}
\begin{split}
e^x&= e^{a1}+(e^{a2}+e^{a3})\cos^2(\piup/3);\\
e^y&= (e^{a2}+e^{a3})\sin^2(\piup/3);\\e^z&= e^{c},
\end{split}
\end{equation}
where $e^{n}=e^{nn}$ is the deformation.
Dislocation loops in each direction are computed as:
\begin{equation}
\begin{split}
\rho^x=& \rho^{a1}+(\rho^{a2}+\rho^{a3})\cos^2(\piup/3);\\
\rho^y=& (\rho^{a2}+\rho^{a3})\sin^2(\piup/3);\\ \rho^z=& \rho^c.
\end{split}
\end{equation}
The corresponding conservation law is 
$ \rho^x+\rho^y=\rho^{a1}+\rho^{a2}+\rho^{a3}.$

In further consideration  we take into account the following  relations between bias coefficients \cite{SG77,RRT,Walgraef}:
$Z^j_{n\,\text{GB}}\gg Z_{\text{iI}}^m\gtrsim Z_{\text{iV}}^j>Z_\text{vI}^m\gtrsim Z_\text{vV}^j>Z_{\text{iN}}^j\gtrsim  Z_\text{vN}^j$, $Z_{n\,\text{GB}}^j=Z_{n\,\text{GB}}$,  $Z_{\text{iI}}^c=Z_\text{vI}^c=0$ (no interstitial loops are formed in the basal plane). It will be shown below that such sink strength difference is responsible for the growth of loops with different rates.  Strict definitions of bias coefficients are based on material parameters  of the alloy (Poisson ratio, shear modulus, loop capture  radii,  grain size, etc.). In most of theoretical and experimental observations it was shown that the values of biased coefficients can be chosen as constants to make an adequate  interpretation of experimental data and predict the material behaviour under irradiation (see, for example, \cite{Was,RRT,Walgraef} and citations therein). On the other hand, it was shown theoretically and experimentally that bias coefficients can vary during the sink size growth at the irradiation that can affect the dynamics of dislocation loops, and, therefore, the radiation growth and swelling. Moreover, the dependence of  bias factors \emph{versus} loop radii becomes  very important at the stage of the growth of small loops. At the stage of their nucleation, one can put constant values of bias coefficients.    By considering grain boundaries, it was shown that $Z_\text{GB}$ should be taken into account as a function of dislocation loop radii to describe the radiation growth and creep of  polycrystalline zirconium alloy \cite{PTG17}. As far as $\alpha{-}\beta$ interface  can be considered as physically identical to the grain boundary, one can assume the same effect of interfaces as the grain boundaries have.  Therefore, the appropriate model allowing one to quantitatively describe the defect structure evolution and the change of mechanical properties in irradiated crystalline systems should not lack a detailed description of irradiation effects even at the level of mean field theory.  In this regard, below we take into account that, at least, bias factors of dislocation loops (of vacancy and interstitial types), grain boundaries and interfaces  are functions of dislocation loop radii growth during irradiation, whereas bias coefficients of network dislocations remain constant. 

In the considered model we put $Z_\text{vN}^j=1$, $Z_{\text{iN}}^j=1.1$ \cite{15,CB05}.
For bias coefficient of dislocation loops we take \cite{RRT}:
\begin{equation}
\begin{split}
Z_{\mu,\Upsilon}^m =\frac{2\piup}{\ln(8 R^m_{\Upsilon}/r_{0\,\mu\Upsilon})}\,,\quad 
Z_{\mu \text{V}}^c =\frac{2\piup}{\ln(8 R^c_\text{V}/r_{0\,\mu \text{V}})};\quad \mu=\{\text{i,v}\},\quad \Upsilon=\{\text{I,V}\}.
\end{split}
\end{equation} 
To satisfy the relation between bias coefficients, one should put $r_\text{0,iI}\gtrsim r_\text{0,iV}>r_\text{0,vI}\gtrsim r_\text{0,vV}$. The loop capture radii, usually, are of the corresponding value of Burgers vectors.
As far as $Z_{\mu,\Upsilon}^j$ are decaying functions \emph{versus} the corresponding loop radius, next  we make limitation of their values by introducing  $Z_{\text{iI}}^{a,\text{min}}$, $Z_{\text{iV}}^{a,\text{min}}$,  $Z_\text{vI}^{a,\text{min}}$,
$Z_\text{vV}^{a,c,\text{min}}$.

Bias coefficient $Z^n_\text{GB}$ is taken according to the definition  \cite{SEZG02,RRT,PTG17}:  
$$
Z_{n\,\text{GB}}=\frac{3\beta_n^2[\beta_n\coth(\beta_n)-1]}{\beta_n^2-3[\beta_n\coth(\beta_n)-1]}\,,\quad \beta_n=\sqrt{\sum_j\rho^j}\lambda_{n\,\text{GB}}\,,\quad \lambda_{n\,\text{GB}}=\lambda_\text{GB}.$$
As was pointed out above, it is natural to assume that $Z_{n\,\text{p}}$ has the same structure as $Z_{n\,\text{GB}}$, where instead of $\lambda_\text{GB}$ one  puts $r_\text{p}$.

In our further consideration we assume that all $m$ directions are equivalent, which allows one to put $\gamma_\text{SF}^{a1}=\gamma_\text{SF}^{a2}=\gamma_\text{SF}^{a3}$, $R_\text{i}^a=R_\text{i}^m$, $R_\text{v}^a=R_\text{V}^m$ and study interstitial and vacancy loops growth in one prismatic plane while the vacancy loops in basal plane.

\section{Simulation and analysis of the system dynamics} \label{sec3}

In order to study  the system dynamics we  use material constants  shown in table~\ref{tab1}.   As far as concentration of Nb is too small and most of Nb is dissolved in Zr matrix (except $\beta$-precipitates) we can  put $D_\text{v}(x)\simeq D_\text{v}^\text{Zr}$ and $D_\text{i}(x)\simeq D_\text{i}^\text{Zr}$. The  interstitials diffusivity $D_\text{i}^\text{Zr}$ is taken in a way to satisfy experimental data at $T=600$~K $D_\text{i}^\text{Zr}\approx8\cdot 10^{-9}$~m$^2$/s at interstitial migration energy $0.2$~eV \cite{koenig1}. 
To calculate the damage rate we take into account that for neutrons of the energy $>1$~MeV bombarding zirconium based alloys ($\sigma_\text{d}=0.184$~barn) one has $\nu_\text{FP}\approx 220$ \cite{koenig1}, at neutron flux $10^{16}$~n/cm$^2$s one gets $K_\text{NRT}\approx 0.4\cdot10^{-6}$~dpa/s. In further consideration,  we put the dose rate equal to $10^{-6}$~dpa/s as a typical value of damage rate for most of reactors.

\begin{table}[!t]\caption{Material parameters used in simulation. \label{tab1}}
\centering
\vspace{2ex}
\begin{tabular}{c|c|c|c}
\hline\hline
Parameter & Value & Dimension& Reference\\
\hline\hline
Lattice parameters for Zr ($a$,$c$)& $(3.2, 5.14)\cdot 10^{-8}$& cm&\\
Atomic volume ($\Omega$)& $3.32\cdot 10^{-23}$& cm$^{3}$&\\
Debye frequency ($\omega_\text{D}$)& $3.82\cdot 10^{13}$& s$^{-1}$&\\ 
Shear modulus ($\mu$)& 33& GPa&\\
Poisson ratio ($\nu$)& 0.34& &\\
Equilibrium vacancy concentration ($c_\text{v0}$) & $0.54\re^{-1.8 \text{eV}/k_\text{B}T}$& at. fract.& \cite{45}\\
Stacking fault energy ($\gamma_\text{SF}^a$, $\gamma_\text{SF}^c$)  &$9.05\cdot 10^{13}$, $1.242\cdot 10^{14}$& eV/cm$^2$ & \cite{China}\\
Thermal  diffusivity of vacancies ($D_\text{v}^\text{Zr}$)  & $1.67\cdot 
10^{-8} \re^{-1.37 \text{eV}/k_\text{B}T}$&  cm$^{2}$/s & \cite{45}\\
Thermal  diffusivity of interstitials ($D_{\text{i}}^\text{Zr}$)& $a^2\omega_\text{D} \re^{-0.2 \text{eV}/k_\text{B}T}$& cm$^{2}$/s&\cite{koenig1} \\
Recombination radius ($r_c$)& $0.62\cdot 10^{-7}$& cm& this work\\
Defect cluster size ($r_{\text{i\,cl}}$, $r_\text{v\,cl}$)& $(0.65$, $0.62)\cdot10^{-7}$& cm& this work\\
Loop capture radius &&&\\
($r_{0\,\text{iI}}$, $r_{0\,\text{iV}}$, $r_{0\,\text{vI}}$, $r_{0\,\text{vV}}$)& $(0.5$,  $0.35$, $0.35$, $0.35)\cdot 10^{-7}$& cm& this work\\
Recombination efficiency $\varepsilon_\text{r}$    & 0.97& & \cite{PTG17}\\
Efficiency of interstitials clustering  $\varepsilon_\text{i}$    &  0.17& & \cite{BGS15}\\
Efficiency of vacancy clustering  $\varepsilon_\text{v}$    &  0.15& & this work\\
Maximal loop number density &&&\\
 ($N^{a,\text{max}}_\text{I,V}$, $N_\text{V}^{c,\text{max}}$)& $ 10^{16}$, $ 10^{15}$ &cm$^{-3}$&\cite{BGS15}\\
Critical dose for $c$-type loops $\phi_0^{c}$&$ 3$&dpa&\cite{BGS15}\\
Terminal dose for  loops nucleation &&&\\($\phi_\text{max}^{a}$, $\phi_\text{max}^{c}$)&$3.84$, $23 $&dpa&\cite{BGS15}\\
Fitting parameter ($A$)&$5 $& &\cite{BGS15}\\
Bias factors of network dislocations &&&\\
 ($Z_\text{vN}^j$, $Z_{\text{iN}}^j$)& 1.0, 1.1& &\cite{15,CB05}\\
Minimal bias factors&&&\\
 ($Z_{\text{iI}}^{a,\text{min}}$, $Z_{\text{iV}}^{a,\text{min}}$,  $Z_\text{vI}^{a,\text{min}}$,
$Z_\text{vV}^{a,c,\text{min}}$) & 1.2; 1.15; 1.12; 1.1& & this work\\
Dislocation density ($\rho_\text{N}^{a}$, $\rho_\text{N}^{c}$) & $10^{8}$, $0.6\cdot10^{8}$& cm$^{-2}$& \\
Displacement rate ($\mathcal{K}_\text{NRT}$) &$ 10^{-6}$& dpa/s& \\
Irradiation temperature ($T$) &$500{-}600$& K &\\
Niobium concentration ($x$) &$0.001{-}0.025$& (at. frac.)&\\
Grain radius ($\lambda_\text{GB}$) & $(5{-}10)\cdot 10^{-4}$& cm &\\
$\beta$-Nb precipitate size ($r_\text{p}$)& $(5{-}10)\cdot 10^{-7}$& cm& \\
Distance from the grain boundary&&&\\
 ($\delta\equiv\ell/\lambda_\text{GB}$)& $0.01{-}1$& &\\
\hline\hline
\end{tabular}
\end{table}

\begin{figure}[!t]
\centering
a)\includegraphics[width=0.45\textwidth]{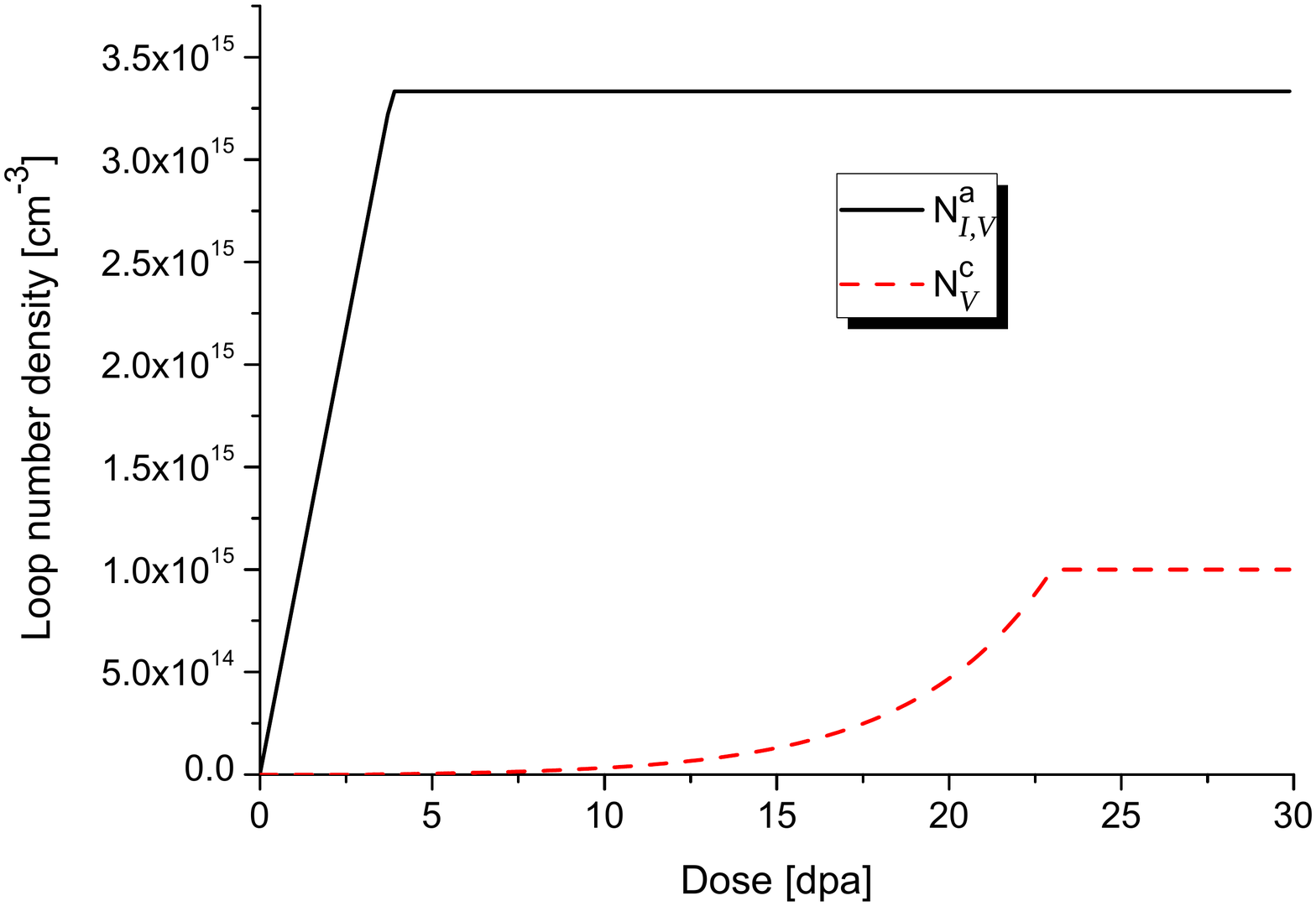}
b) \includegraphics[width=0.45\textwidth]{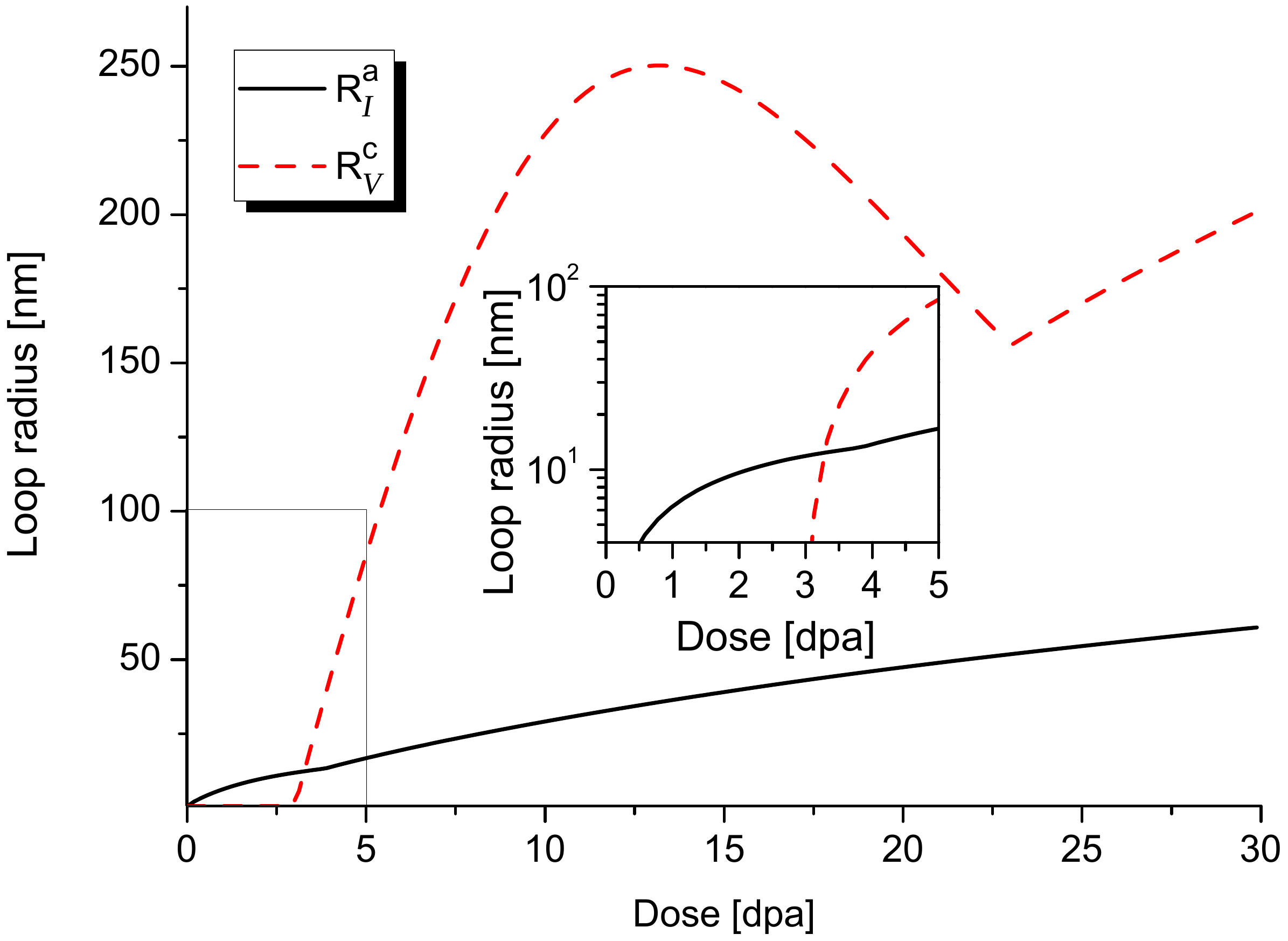}\\ 
c) \includegraphics[width=0.45\textwidth]{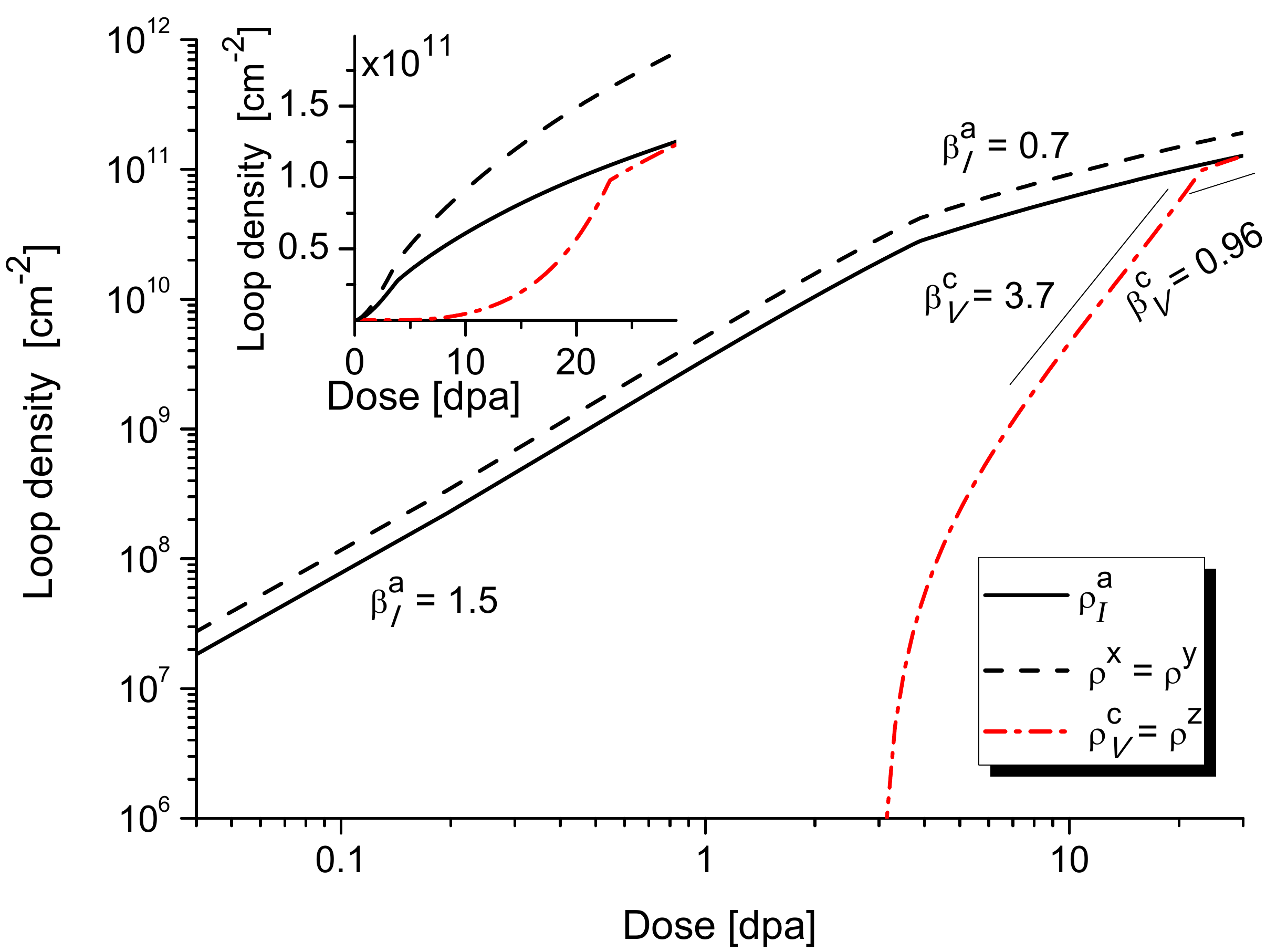}
d) \includegraphics[width=0.45\textwidth]{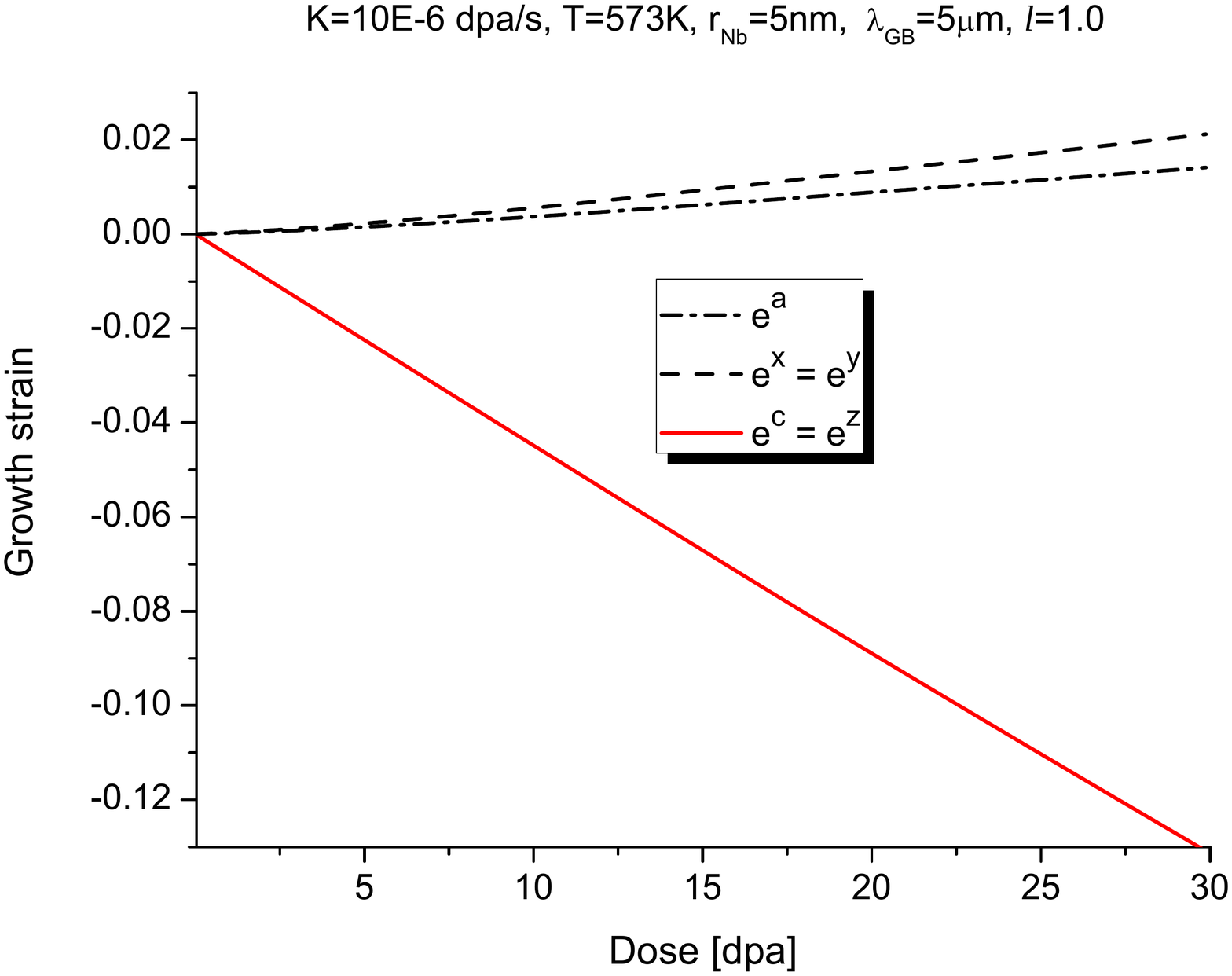} 
\caption{(Colour online) Dose dependencies of loop number densities (a), radii of $\langle a\rangle$-type and $\langle c\rangle$-type loops (b), loop densities (c), and growth strains (d) in the center of the grain at $x=2.5\%$, $\mathcal{K}=10^{-6}$~dpa, $T=573$~K. Other parameters are:  $r_\text{p}=5$~nm, $\lambda_\text{GB}=5$~{\textmu}m. \label{fig1}}
\end{figure}

We start our study by considering the alloy Zr-2.5\%Nb as one of the main alloys used for cladding materials. 
Dose dependencies of  loop number densities are shown in  figure~\ref{fig1}~(a), where kinks  in curves relate to the change in speeds of  loops nucleation as was shown by equations~(\ref{Na}), (\ref{Nc}). The corresponding values are independent of concentration of Nb. 
In figure~\ref{fig1}~(b) we plot the dependencies of radii of loops formed in these alloys during irradiation. It follows that only interstitial $\langle a\rangle$-type loops emerge and grow constantly. The $\langle a\rangle$-type vacancy   loops are unstable and do not grow independently of initial conditions, whereas 
$\langle c\rangle$-type vacancy   loops grow from the critical dose $\phi_0^c$. Their radius grows faster than the radius of interstitial loops. By considering the corresponding sizes at small doses [see insertion in figure~\ref{fig1}~(b)], one finds that at doses up to $5$~dpa, interstitial loop radius  grows from several nanometers up to $10$~nm, whereas $\langle a\rangle$-type loops critically  attain the radius of $\lesssim10$~nm which  grows up to $70$~nm. Our results at small doses relate well to the theoretical work \cite{BGS15}, where the dynamics of the loop number density was not included into the loop radii dynamics, and are consistent with experimental observations discussed in the review of experimental studies \cite{koenig1}.   At large doses one gets a continuous growth of the radius of interstitial $\langle a\rangle$-type loops up to $50$~nm, whereas  $\langle c\rangle$-type loop radius manifests non-monotonous behaviour caused by non-linear dynamics of their number density $N_\text{V}^c$ limiting their growth rate. The same effect was observed in \cite{PTG17}, where the corresponding terms were considered in the model of the loop radii growth even with the model for loop  nucleation according to experimental data from  SIRIUS reactor.    The size of $\langle c\rangle$-type loops is much larger  than interstitial loops have which increases up to $250$~nm,  whereas their number density is small compared to interstitial loops. 

It is easy to see that the indicated non-monotonous dynamics of $R_\text{v}^c$ is related to the dynamics of the loop number density that gives different regimes of the loop density evolution [see figure~\ref{fig1}~(c)]. These regimes are limited  by the terminal loop number density $N_\text{I}^{a,c,\text{max}}$. It is seen that the loop densities  attain the values three orders larger than the network dislocation densities $\rho^a_\text{N}\approx 10^{8}$~cm. Formally, both $\langle a\rangle$ and  $\langle c\rangle$-type loops grow with two regimes and take  practically the same values at $\phi>\phi_\text{max}^c$.   These dose regimes can be approximated by power-law dependencies: $\rho_\text{I}^a\propto \phi^{\,\beta_\text{I}^a}$ and $\rho_\text{V}^c\propto \phi^{\,\beta_\text{V}^c}$, where growth exponents $\beta_\text{I}^a$ and $\beta_\text{V}^c$  slightly depend on the grain size, distance from the grain boundary, size of precipitates and irradiation conditions  (dose rate and temperature). For the special choice $\mathcal{K}=10^{-6}$~dpa/s and  $T=573$~K one has $\beta_\text{I}^a\approx1.5$ (for doses $\phi<3$~dpa) and $\beta_\text{I}^a\approx0.7$ at large doses.  It follows that at large doses ($\phi>\phi_\text{max}^c$) the dynamics of $\langle c\rangle$-type loops is characterized by the linear law  with $\beta_\text{V}^c\approx 1$. At small doses one gets a complex behaviour of $\rho_\text{V}^c$ due to the complicated  dynamics of $N_\text{V}^c$ and $R_\text{V}^c$. At elevated doses ($15<\phi<23$~dpa) one gets  $\beta_{\text{V},r}^c\approx3.7$; at extremely large doses ($\phi>23$~dpa) the linear regime $\rho_\text{V}^c\propto \phi$ is observed.
The obtained values for loop densities relates well with experimental observations for zirconium alloys irradiated by neutrons in the reactor SIRIUS (see discussions in \cite{PTG17}). 

The dynamics of growth strains is shown in figure~\ref{fig1}~(d). Here, a weak linear growth of $e^x=e^y$ with a fast  decrease of the strain $e^z$ is observed.
It is  caused by the presence of grain boundaries having the largest bias factor compared to the other sinks. At large grains [$\lambda_\text{GB}\sim (40{-}50)$~{\textmu}m] deformations $e^z$  become smaller, whereas $e^x=e^y$ will be larger due to a small contribution of grain boundaries, where the main sinks of point defects are network dislocations and dislocation loops (the effect of $\beta$-phase precipitates becomes small too, due to their location at grain boundaries).

By considering the number of defects in the loops calculated by the formula $\piup (R^{\,j}_\text{I,V})^2b^{\,j}/\Omega$, where $\Omega_a$ is the atomic volume, one finds that the number of defects in interstitial loops varies in the interval $10^2{-}10^3$ at doses  up to $5$~dpa.  For $\langle c\rangle$-type loops,  the number of vacancies in loops is around $10^2{-}10^5$ at doses from 3 up to $5$~dpa.

\begin{figure}[!b]
\centering
a)\includegraphics[width=0.45\textwidth]{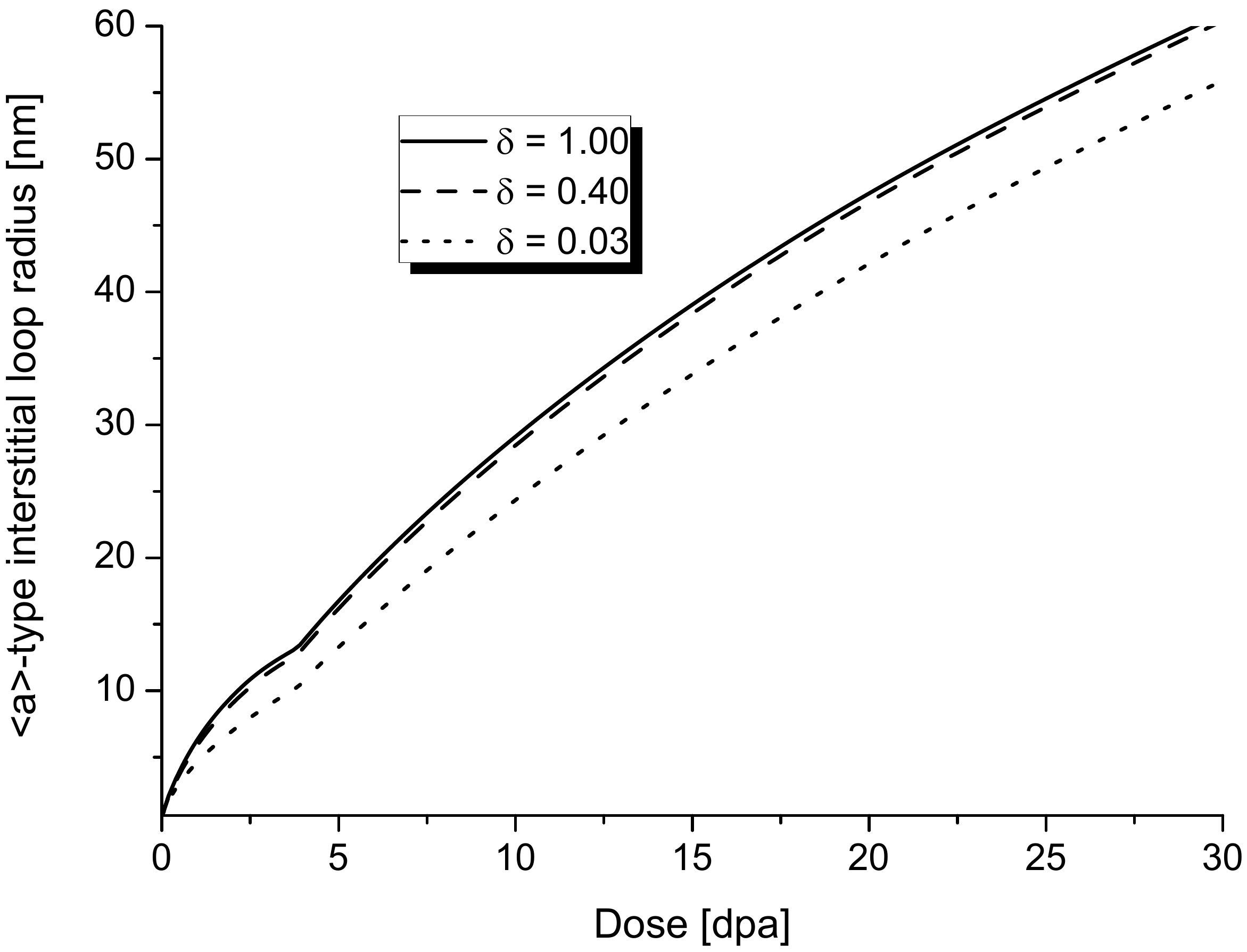}
b) \includegraphics[width=0.45\textwidth]{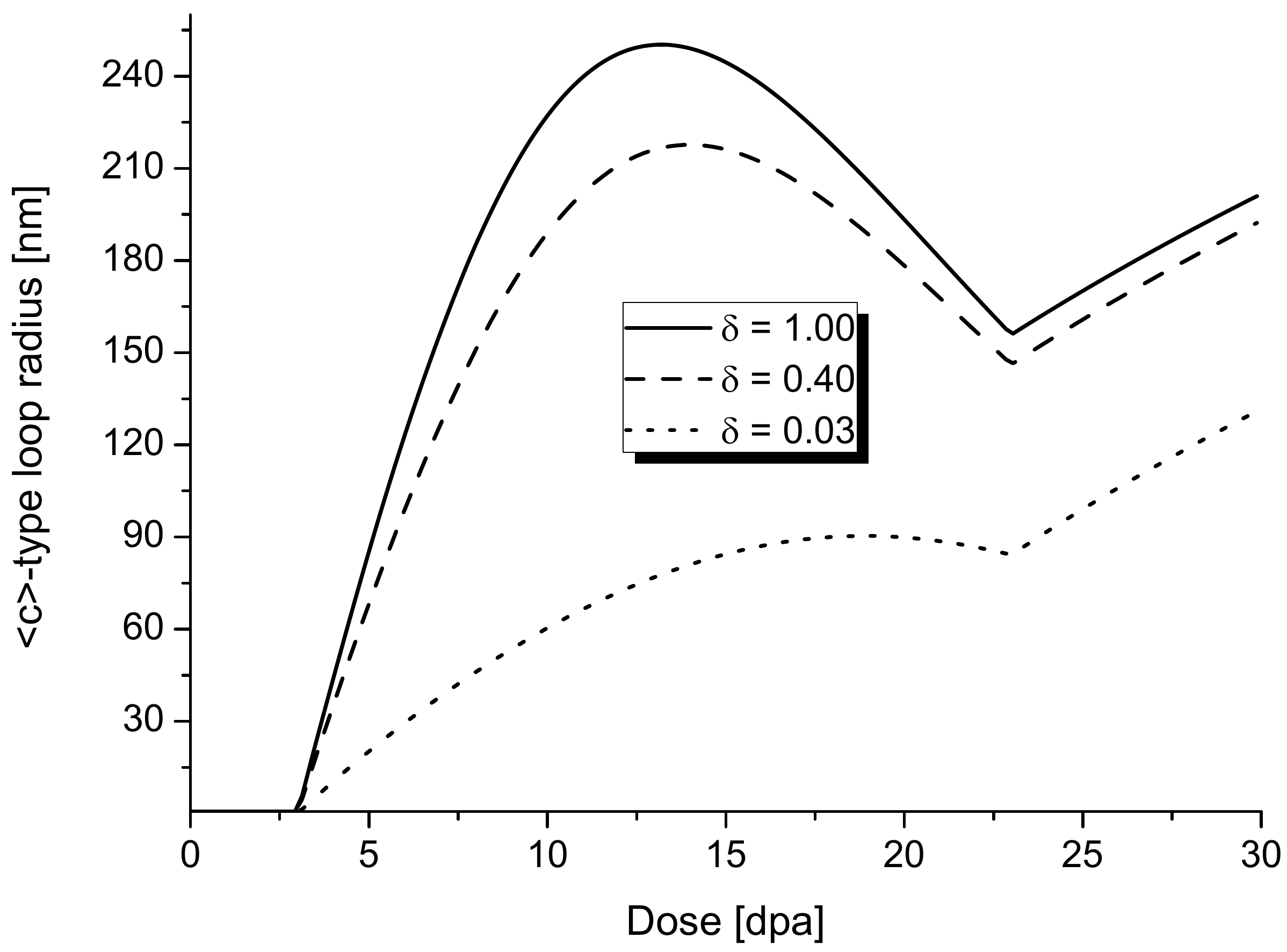}\\
c) \includegraphics[width=0.45\textwidth]{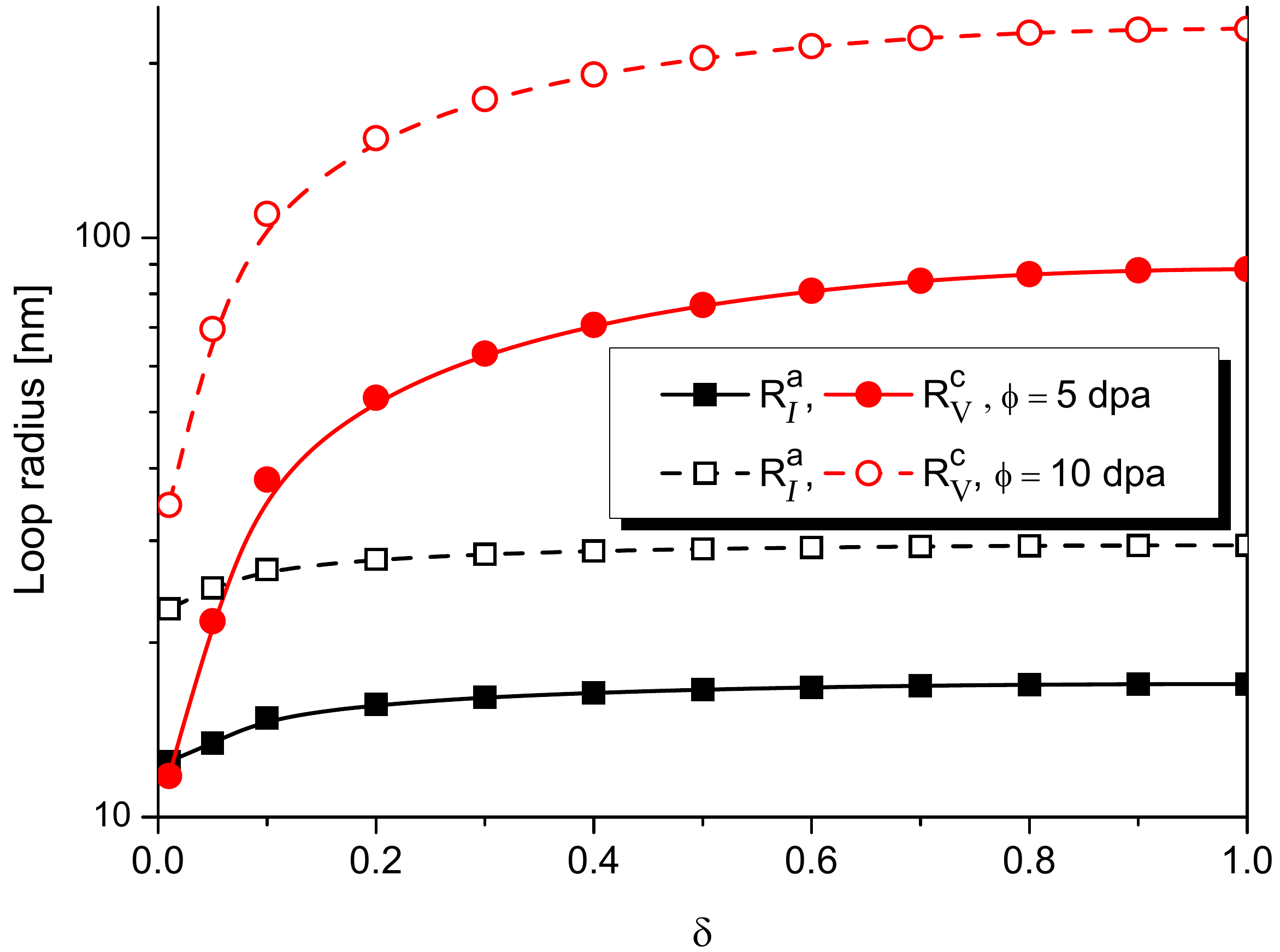}
d) \includegraphics[width=0.45\textwidth]{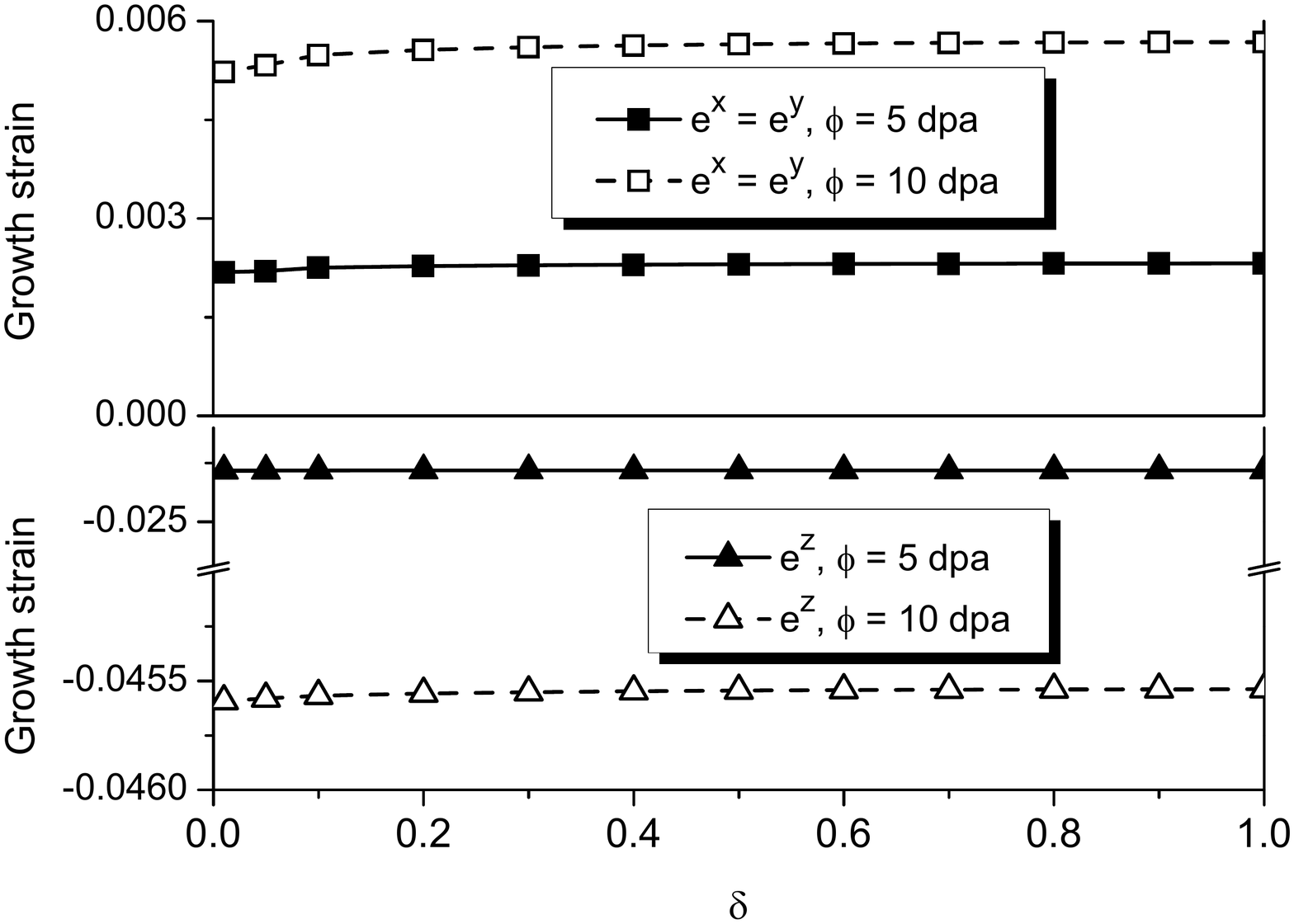}
\caption{(Colour online)  Dose dependencies of interstitial loop radius (a) and vacancy loop  radius (b) at different location from the grain boundary. (c) Loop radii and growth strain (d) dependencies \emph{versus} distance from the grain boundary  at doses 5 and 10 dpa. Calculations were done  at  $\mathcal{K}=10^{-6}$~dpa, $T=573$~K,  $r_\text{p}=5$~nm, $\lambda_\text{GB}=5$~{\textmu}m. \label{fig2}}
\end{figure}

To  analyse a change in  material properties  at different locations from the grain boundary we compute the above dependencies at different values of $\delta=\ell/\lambda_\text{GB}$ shown in figure~\ref{fig2}.
By comparing curves of loop radii at different $\delta$ in figure~\ref{fig2}~(a) one finds that interstitial loops located in the vicinity of the grain boundary are of the same size as the loops in the center of the grain at small doses, when loops grow from clusters of point defects. The loop size decreases   down to $5$~nm in the  vicinity of the grain boundary at  elevated doses ($\phi>5$~dpa) compared to the loops growing in the center of the grain. This effect is  explained by a strong  absorption of point defects and their  clusters  by grain boundary and phase interfaces. The most essential changes are observed for sizes of $\langle c\rangle$-type loops at different locations in the grain [see figure~\ref{fig2}~(b)]. Here, the growth rate of loops in the center of the grain  is much larger  than the one for loops far from the center. The size of loops decreases down to $\approx150$~nm. This effect is observed for the doses up to the terminal one, $\phi<\phi_\text{max}^c$. Above  $\phi_\text{max}^c$ the growth rate of loops is the same independently of the location of $\langle c\rangle$-type loops  inside the grain. Here, the difference in the loops radii is about $90$~nm. 
Growth strains do not manifest observable dependencies on the distance from the grain boundary (not shown here). To illustrate the distributions of loops inside the grain we choose the corresponding values from the obtained protocols at two fixed doses [see figure~\ref{fig2}~(c)]. This means that  a distribution  map of interstitial loops inside the grain is uniform except a very short distance from the grain boundary. At the same time,  a distribution of $\langle c\rangle$-type loops is non-uniform. The largest difference in the loop sizes  is observed at distances up to a half of the grain independently of the accumulated dose. A distribution map of growth strain inside the grain is shown in figure~\ref{fig2}~(d). It is uniform at small doses ($\phi=5$~dpa); the difference  is observed for $e^x=e^y$ at elevated doses ($\phi=10$~dpa). Therefore, a deformation of the crystal inside the grain can be considered as non-local.

\begin{figure}[!b]
\centering
a)\includegraphics[width=0.45\textwidth]{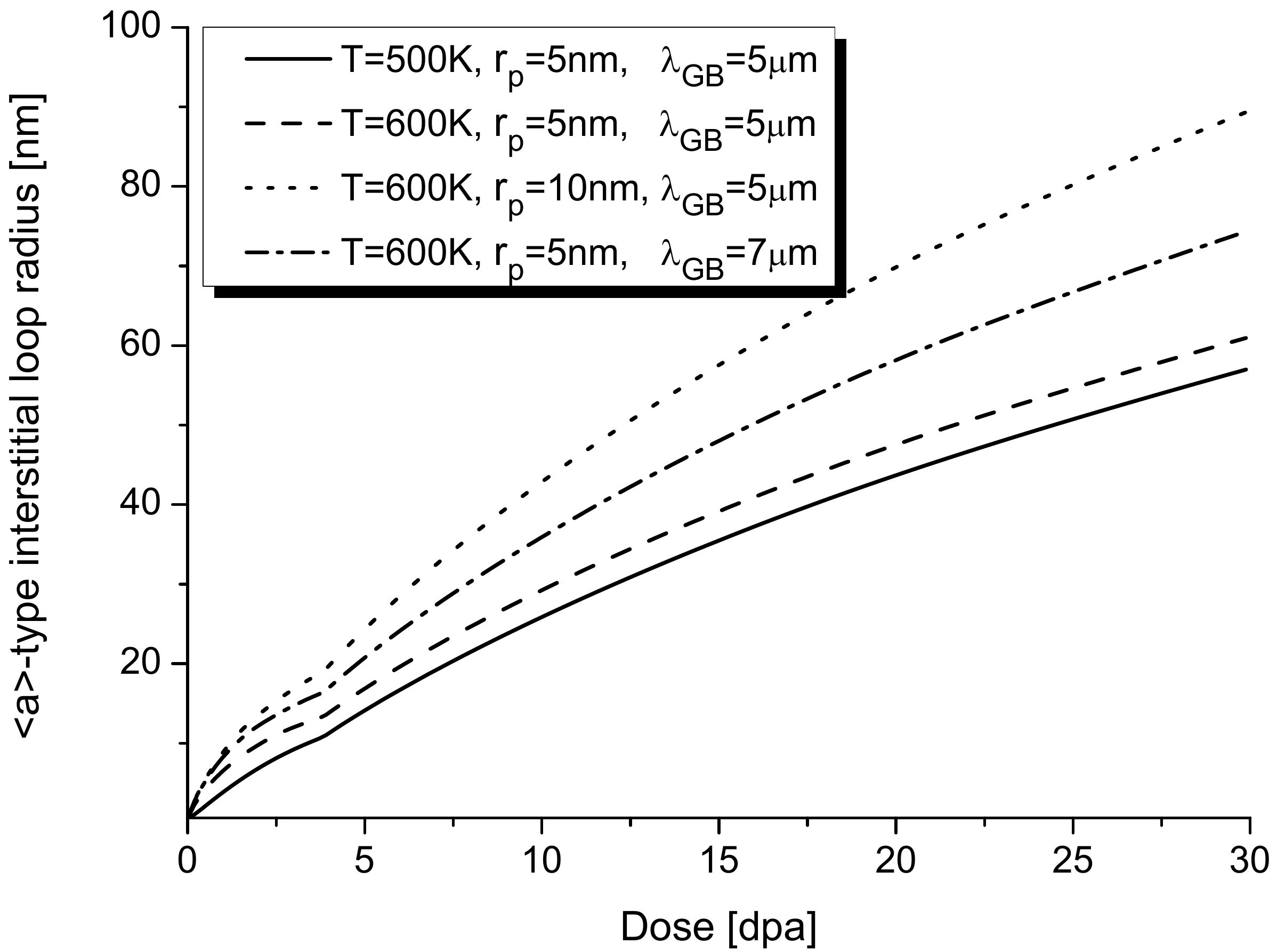}
b) \includegraphics[width=0.45\textwidth]{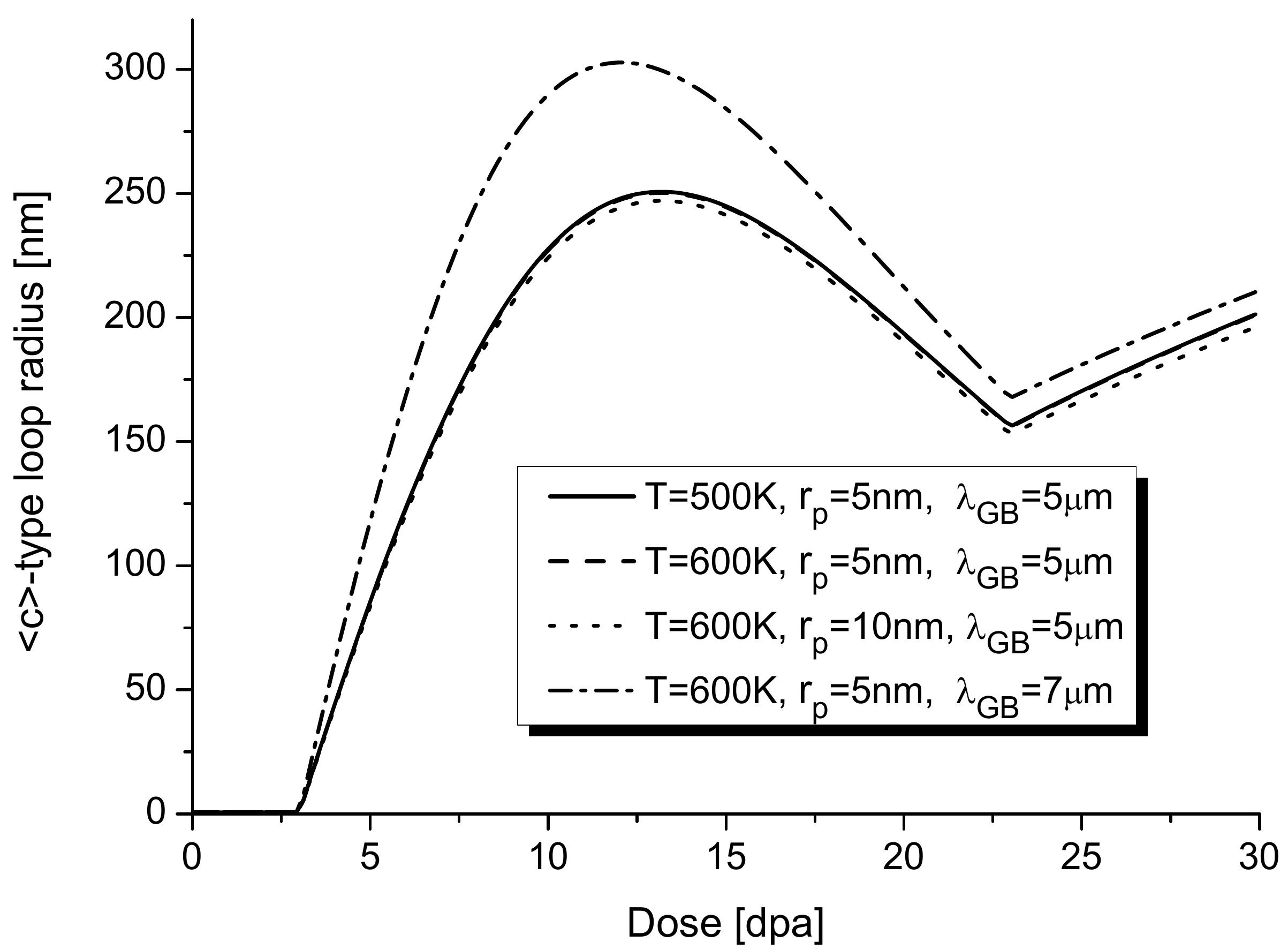}\\
c) \includegraphics[width=0.45\textwidth]{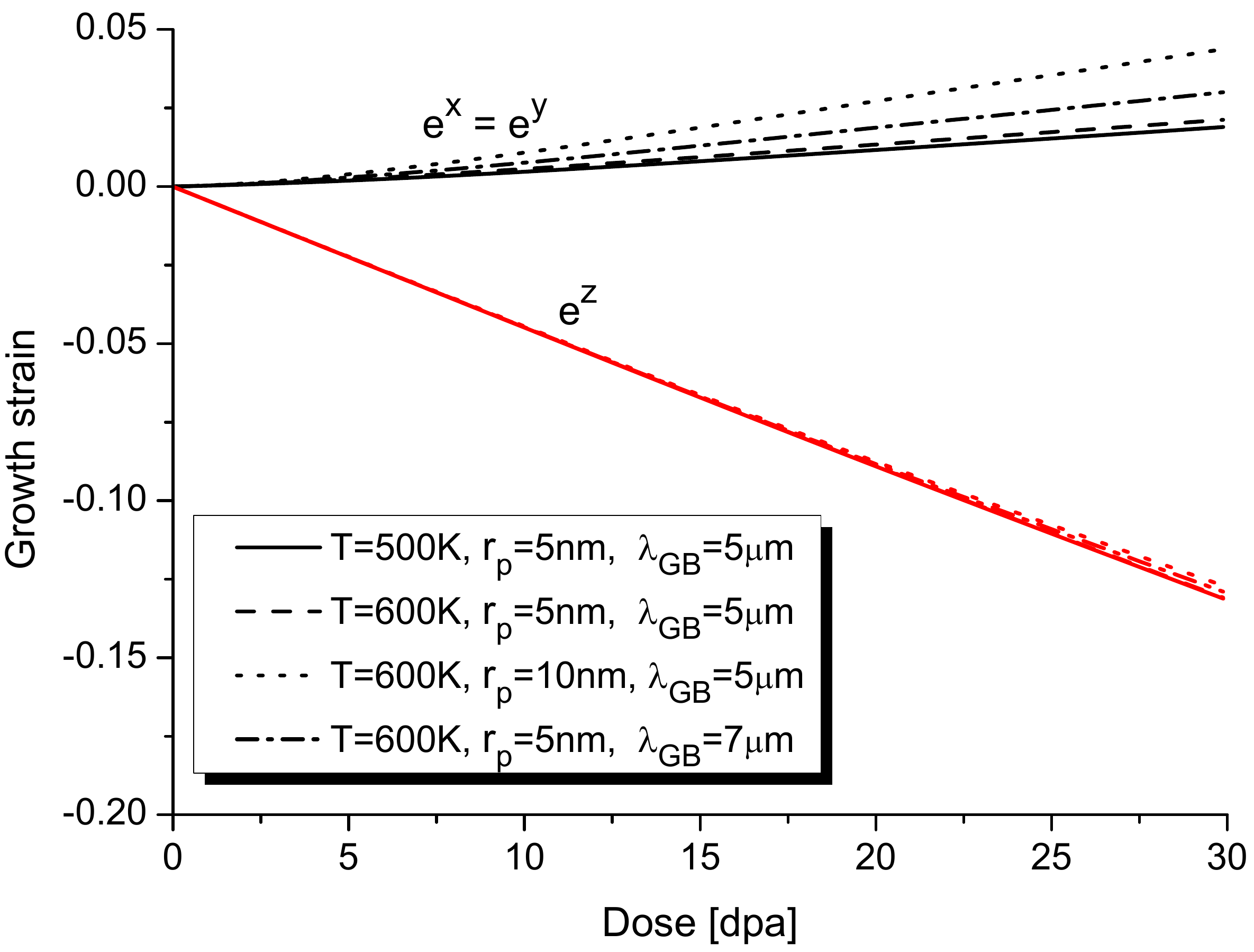}\\
\caption{(Colour online)  Dose dependencies of interstitial loop radius (a),  $\langle c\rangle$-type loop  radius and growth strains (c) at different temperatures, grain size and precipitate size at  $\mathcal{K}=10^{-6}$~dpa and $\delta=1.0$. \label{fig3}}
\end{figure}

The  dose dependencies of interstitial loops  radius at different grain size, precipitate size and temperature  are shown in figure~\ref{fig3}~(a).  
It follows that with the temperature growth the size of loops increases, loops grow at the same growth rates (cf. solid and dash curves at different irradiation temperature).  An increase in the temperature results in a  growth of the net flux of defects to dislocation loops.   An increase in the grain size (cf. dash and dash-dot lines)  promotes a faster growth of the loop radii. At large grain sizes, most of point defect clusters survive  inside large grains promoting the growth of the loops radii and here the point defects relax mostly on network dislocations. At small grain sizes,  most of clusters and point defects are absorbed by  grain boundaries as main sinks.  The same effect is observed when the size of precipitates increases (cf. dot and dash curves). As was discussed  previously, phase interfaces play the same role as the grain boundary, hence the effect of their action is the same (an increase in the precipitate size localized at the grain boundary effectively  increases the grain boundary). By considering the dynamics of $\langle c\rangle$-type loops [see figure~\ref{fig3}~(b)] one finds that a change in the temperature from $500$~K up to $600$~K does not lead to essential changes in their  radius (here, solid and dash curves  coincide). A bit different situation is observed for $\langle c\rangle$-type loops radii  when the size of both grain and precipitate increases. Here, with the grain size growth one can observe an increase in values of the vacancy loop radius. At the same time, with the growth in precipitate size the vacancy loop size slightly decreases due to a decrease in  the net flux of point defects to $\langle c\rangle$-type loops. By considering the dose dependencies of growth strains in figure~\ref{fig3}~(c) one can conclude that with an increase in the temperature the quantities $e^x=e^y$ grow slightly and a small decrease in  $e^z$ is observed. An essential influence on the growth strains  is produced by grain  and/or precipitate sizes. With their growth one gets elevated values of $e^x=e^y$ and small values of  $e^z$ only at large doses. At small doses, no differences in the growth strain values were observed.

\begin{figure}[!b]
\centering
a)\includegraphics[width=0.45\textwidth]{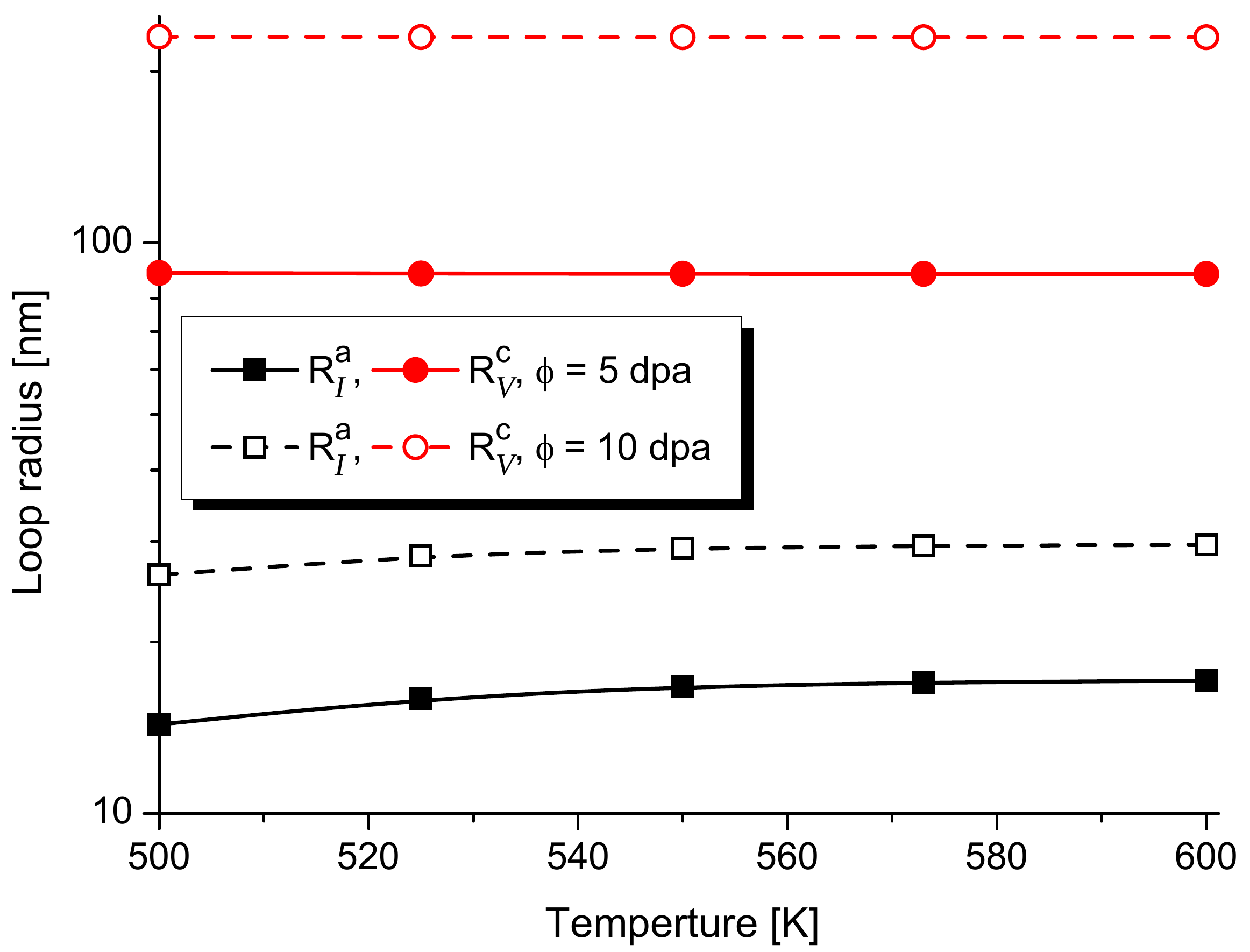}
b)\includegraphics[width=0.45\textwidth]{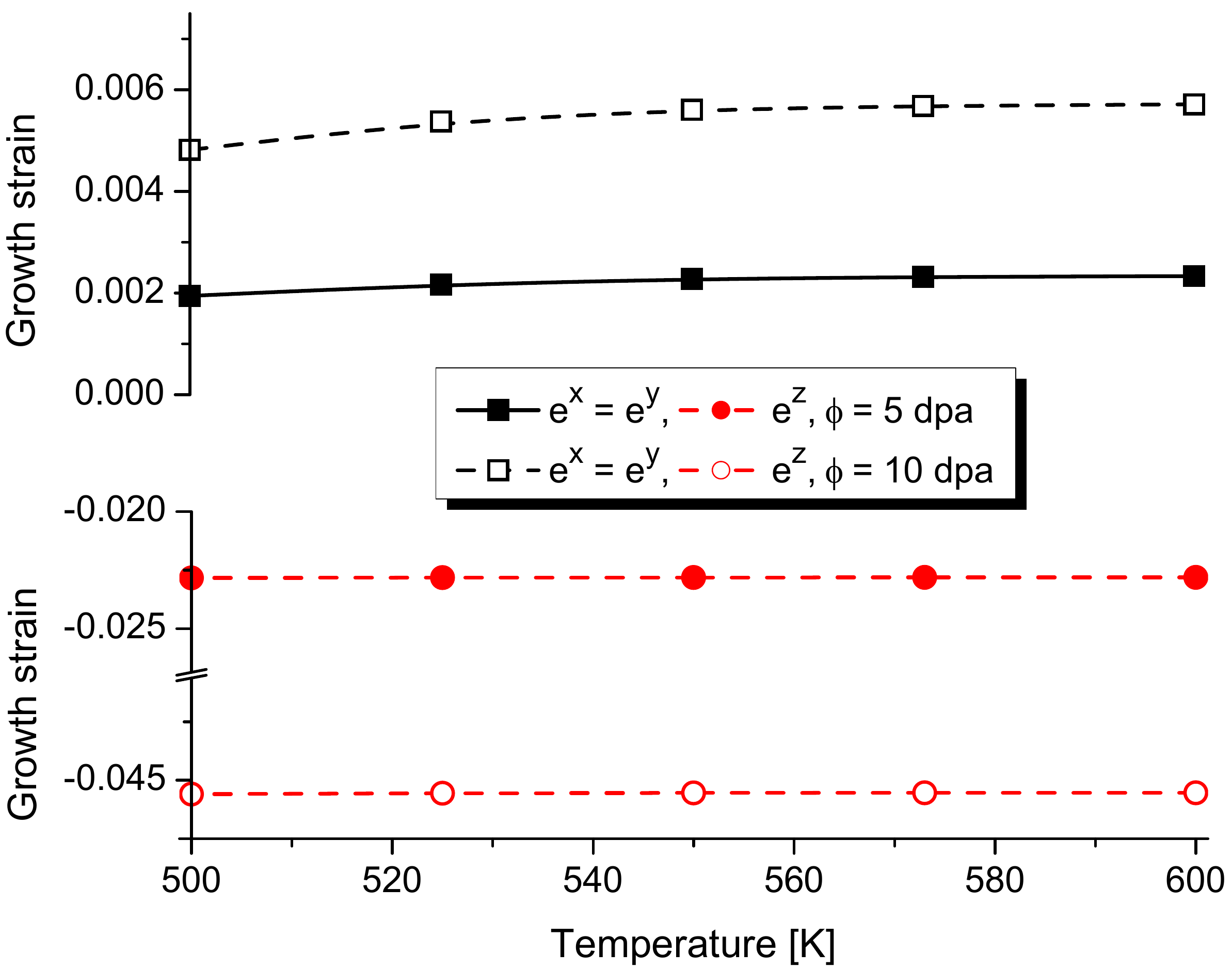}
\caption{(Colour online)  Temperature dependencies  of dislocation loop radii (a) and growth strains (b) at different doses at  $\mathcal{K}=10^{-6}$~dpa,  $\lambda_\text{GB}=5$~{\textmu}m, $r_\text{p}=5$~nm and  $\delta=1.0$. \label{figtmp}}
\end{figure}

To illustrate the temperature influence onto the loop radii and growth strains we choose the data from the obtained protocol at 5 and $10$~dpa shown in figure~\ref{figtmp}. From figure~\ref{figtmp}~(a) it follows that an observable change in values of interstitial loop radius can be found only with the temperature growth from $500$~K up to $550$~K independently of the accumulated doses. The size of $\langle c\rangle$-type loops does not change  with the temperature variation. By considering the growth strains dependencies on temperature one can see  that the values of $e^z$ are  independent of the temperature, whereas slight temperature dependence of $e^x=e^y$ is well seen only at elevated doses ($\phi\simeq 10$~dpa).

\begin{figure}[!t]
\centering
a)\includegraphics[width=0.45\textwidth]{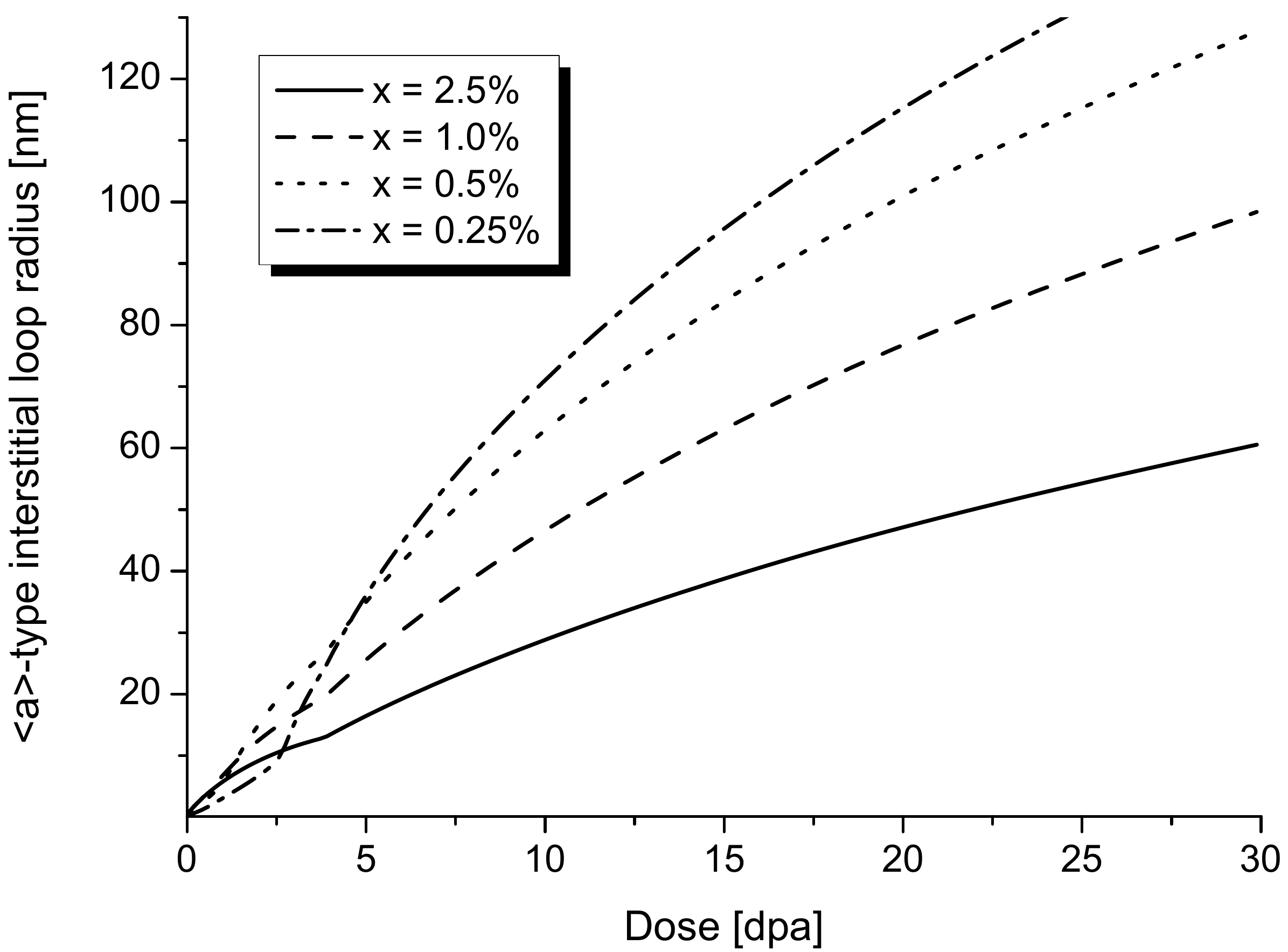}
b)\includegraphics[width=0.45\textwidth]{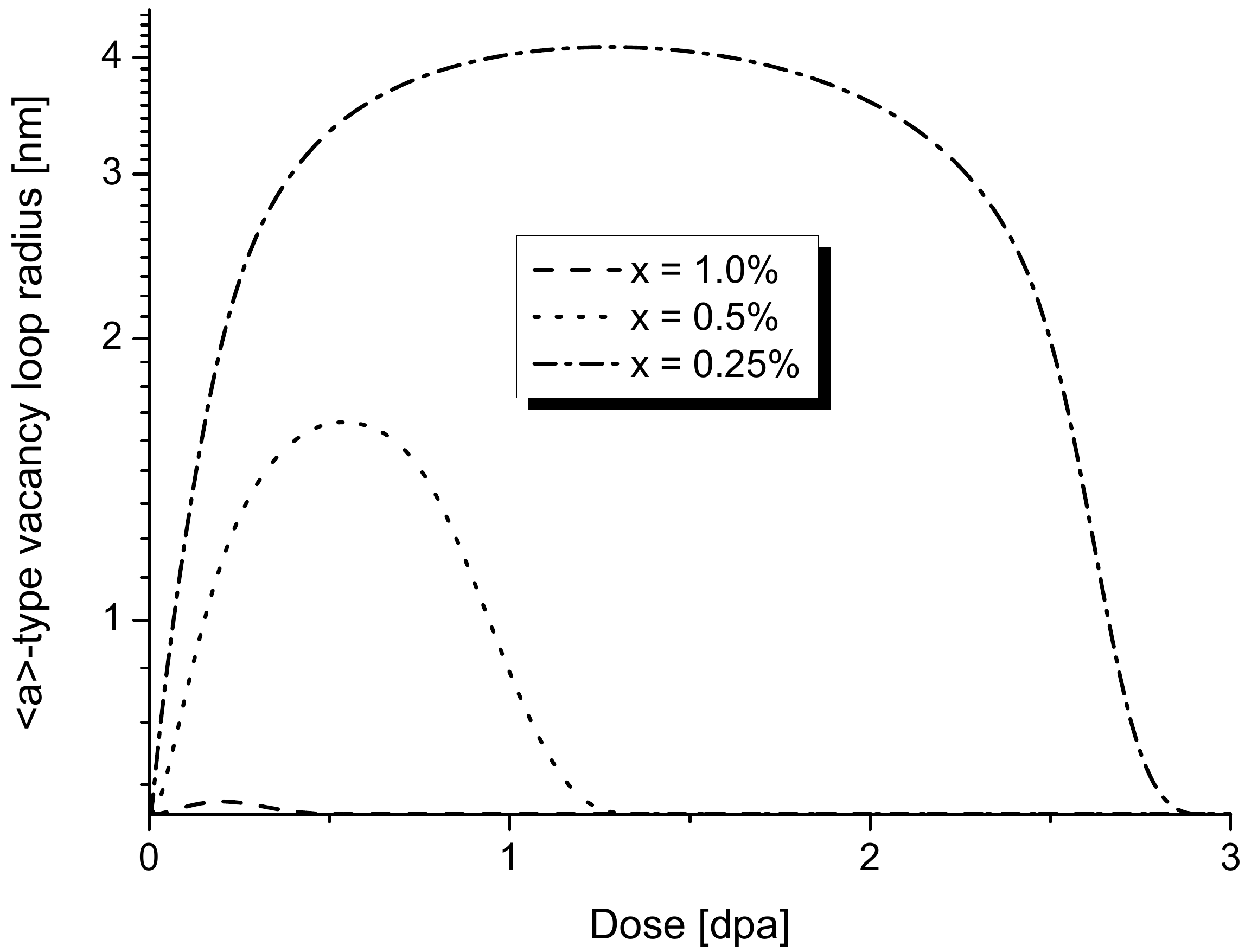}\\
c)\includegraphics[width=0.45\textwidth]{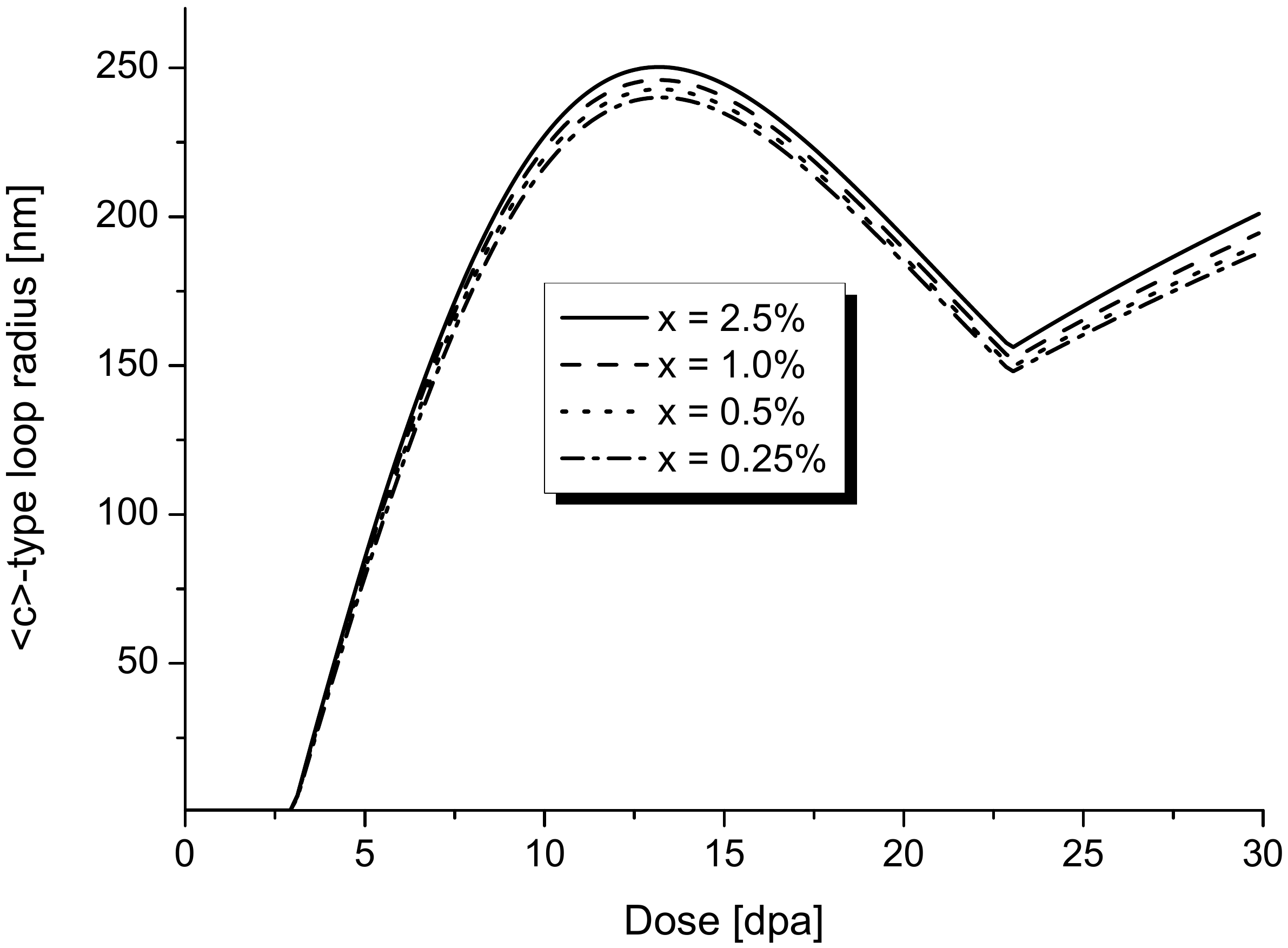}
d)\includegraphics[width=0.45\textwidth]{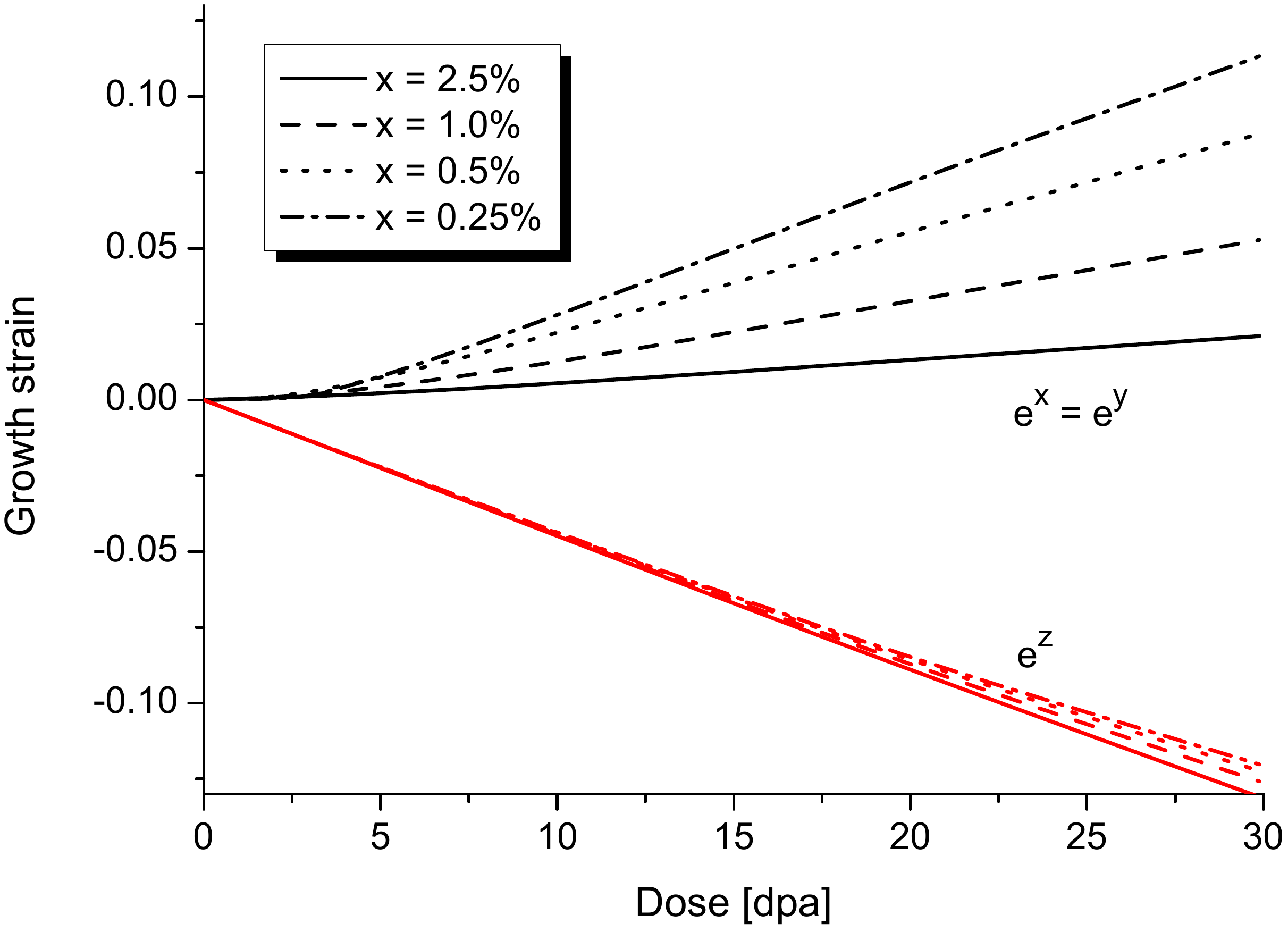}
\caption{(Colour online)  Dose dependencies of interstitial loop radius (a), 
 $\langle a\rangle$-type vacancy  radius (b),
 $\langle c\rangle$-type loop  radius (c) and growth strains (d) at different content of Nb  at  $\mathcal{K}=10^{-6}$~dpa, $T=550$~K, $\lambda_\text{GB}=5$~\textmu m, $r_\text{p}=5$~nm and  $\delta=1.0$. \label{fig4}}
\end{figure}

The above results were addressed to Zr-2.5\%Nb alloy irradiated at reactor conditions. Next we consider the case when the content of niobium varies from 2.5\% down to 0.25\% (see figure~\ref{fig4}). From dose dependencies of interstitial loop radii shown in figure~\ref{fig4}~(a) it follows that with a decrease in niobium concentration the size of interstitial loops increases at elevated doses. At small doses and small Nb content,  their dynamics and size strongly depend on the dynamics and  size of metastable $\langle a\rangle$-type vacancy loops shown in figure~\ref{fig4}~(b). Vacancy $\langle a\rangle$-type loops can be formed only at a small concentration of Nb and/or at low temperatures compared to the typical irradiation conditions  realized (here, we take $T=550$~K). Their size is of several nano-meters and they can be observed at small doses.  It is seen that at small niobium content, a growth of  $\langle a\rangle$-type vacancy loops suppresses the dynamics of  the interstitial loops (cf. dash and dash-dot curves in figure~\ref{fig4}~(a) obtained at $x=1.0\%$ and 0.25\%). At the stage of $\langle a\rangle$-type vacancy loops collapse, the size of interstitial loops manifests a fast growth.  By studying the dynamics of  $\langle c\rangle$-type loops shown in figure~\ref{fig4}~(c) one concludes that with a decrease in Nb content their size slightly decreases with no significant changes in a growing dynamics. The most essential influence of Nb concentration is observed for growth strains at large doses [see figure~\ref{fig4}~(d)]. Here, by comparing curves for $x=2.5\%$ and $x=0.25\%$ one can observe an increase in the values of components $e^x=e^y$ up to 0.1\% at doses $\phi\approx 30$~dpa. The quantity $e^z$ does not manifest essential changes except decreasing its absolute values by a few hundredths of a percent.   

\begin{figure}[!t]
\centering
a)\includegraphics[width=0.45\textwidth]{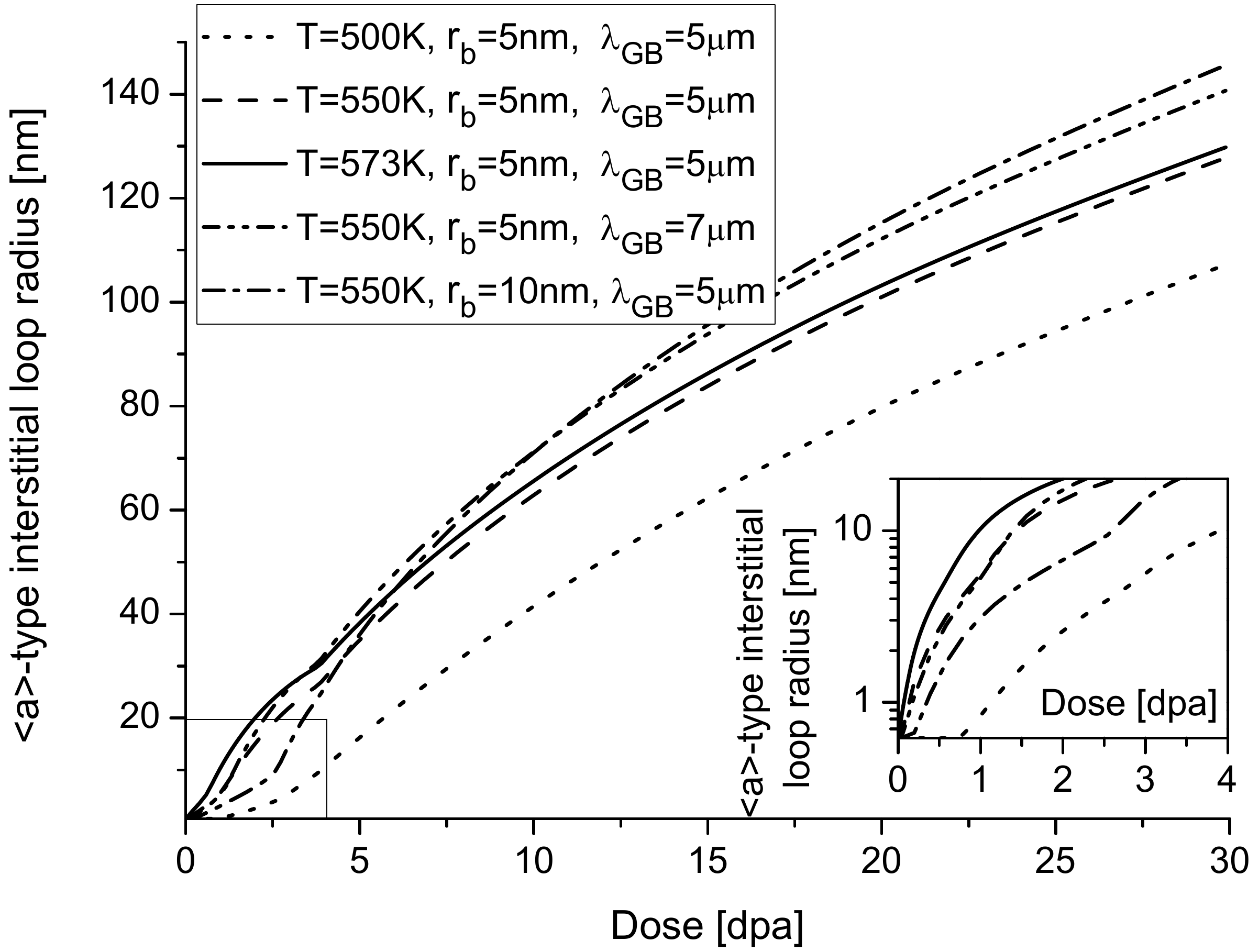}
b)\includegraphics[width=0.45\textwidth]{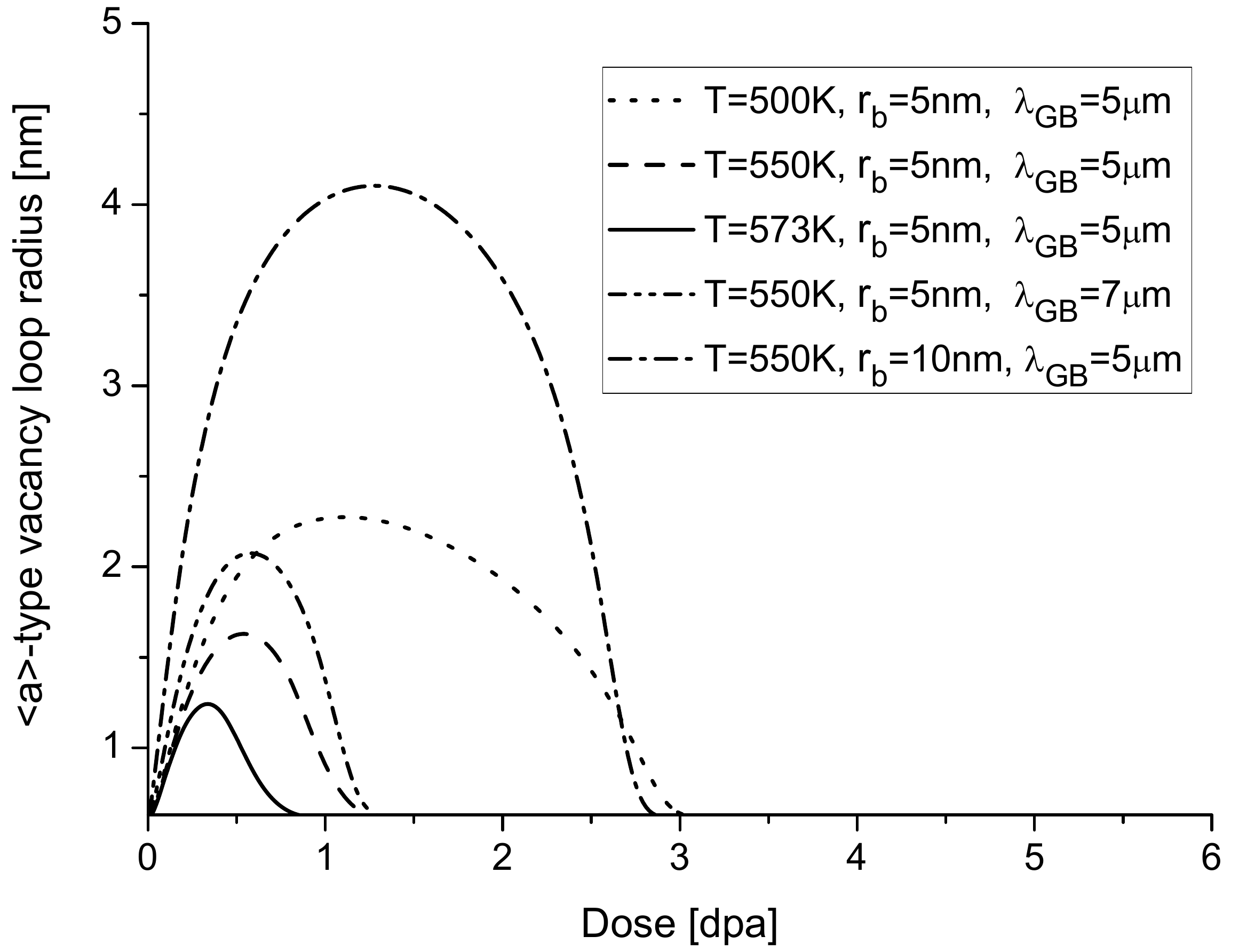}
\caption{ Dose dependencies of  $\langle a\rangle$-type interstitial loop radius (a) and  vacancy  radius (b) at $x=0.5\%$, $\mathcal{K}=10^{-6}$~dpa and  $\delta=1.0$. \label{fig5}}
\end{figure}
\begin{figure}[!t]
\centering
\includegraphics[width=0.45\textwidth]{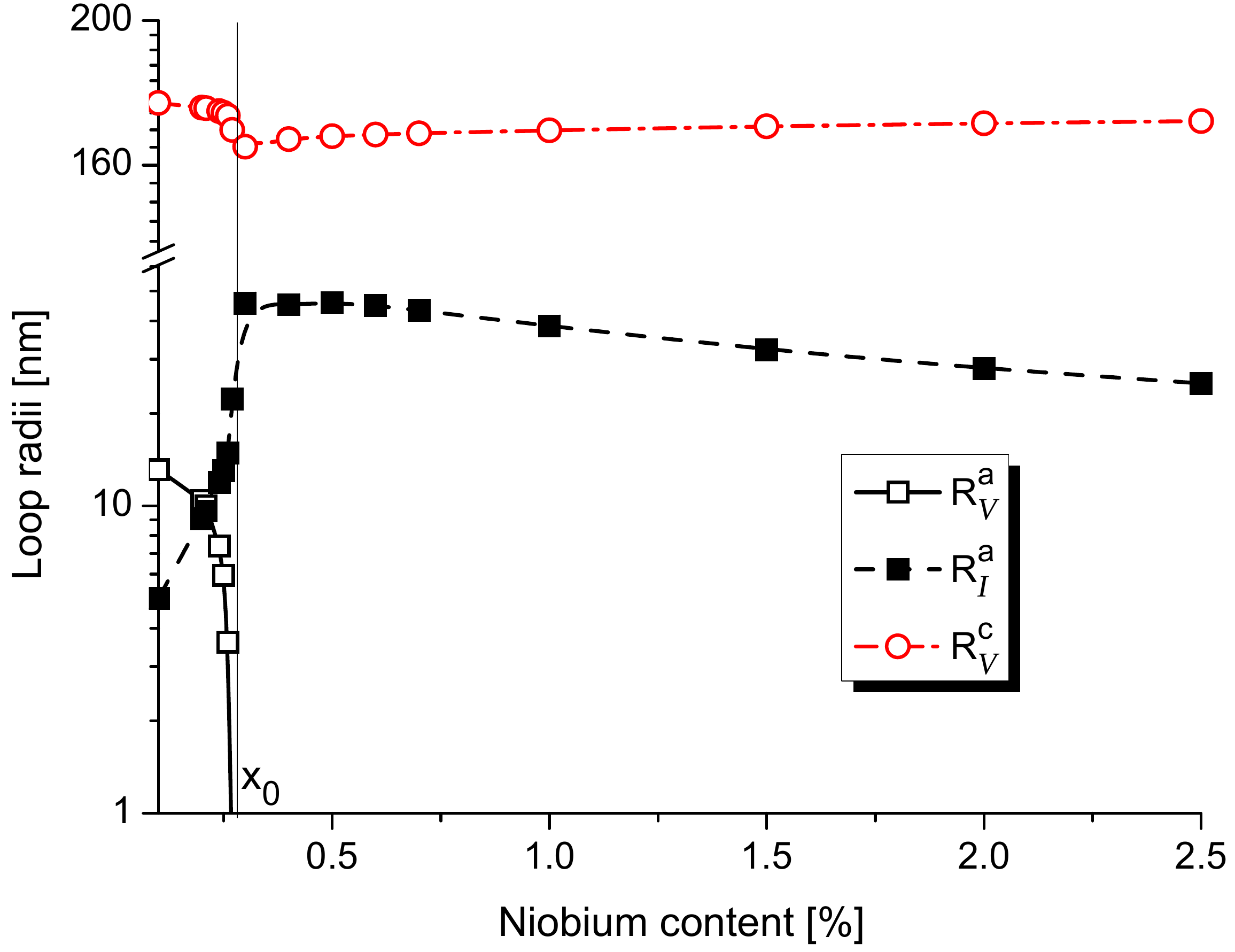}
\caption{(Colour online) Dependencies of the loops radii \emph{versus} niobium content at $T=550$~K, 
$\mathcal{K}=10^{-6}$~dpa, $\phi=5$~dpa, $\lambda_\text{GB}=5$~\textmu m, $r_\text{p}=5$~nm and  $\delta=1.0$. \label{fig7}}
\end{figure}

Next, let us discuss   $\langle a\rangle$-type loops growth at small content of Nb shown in figure~\ref{fig5}. It follows that with the temperature decrease the incubation dose for interstitial loops growth increases, at $T=500$~K interstitial loops (of the size $>1$~nm, experimentally observed by TEM)  grow after $\phi>1 $~dpa.   From figure~\ref{fig5}~(b) it follows that the dynamics and the size of $\langle a\rangle$-type vacancy loops is governed by temperature (discussed above), grain size and precipitate size. Here, with the temperature decrease down to $500$~K the dose interval of vacancy loops existence increases. By clarifying the curves at different grain sizes (dash and dash-dot-dot) one finds that the size of vacancy loops grows up in the grain of large size. With an increase in the precipitate size both  size of vacancy loops of $\langle a\rangle$-type and  dose interval of their existance increase (cf. dash and dash-dot lines).

By using the obtained protocols of loop radii dose dependencies at different concentrations of the niobium, one gets the dependencies of the loop radii \emph{versus} $x$ at a fixed temperature and dose shown in figure~\ref{fig7}. It follows that an increase  in the niobium concentration suppresses the formation of $\left<a\right>$-type vacancy loops (see empty squares) and results in a decrease in $\left<c\right>$-type loop size (see empty circles) at $x<x_0$, where $x_0\approx 0.25\%$ at other fixed parameters. At the same time,  interstitial loop size increases with $x<x_0$ (see filled squares).  Above the threshold  $x_0$ vacancy  $\left<a\right>$-type  loops are unstable and collapse, whereas  $\left<c\right>$-type loops manifest a slight growth with niobium content; the radius of interstitial loops decreases with $x$. From the provided calculations it follows that the threshold value $x_0$ depends on irradiation conditions (temperature and dose rate)  and material parameters (grain and precipitate sizes, network dislocation densities, location out of the grain boundary).

\section{Conclusions} \label{sec4}

We propose a generalized model to study the defect structure evolution in irradiated  binary alloys which takes into account the influence of  vacancy and interstitial  clusters, secondary phase precipitates,  grain size  and concentration of alloying elements onto the dynamics of dislocation loops growth. 
The developed model allows one to  study the locality of  loops growth inside grains by considering point defect recombination and different bias  factors of different sinks for point defects and their clusters at different irradiation conditions (temperature and dose rate).
 This model is exploited to describe dislocation loops growth and the associated radiation growth in zirconium-niobium  alloys under neutron irradiation. 

By studying Zr-2.5\%Nb alloy at reactor conditions it was found that growing $\langle a\rangle$-type loops are only of interstitial character. By considering the dynamics of interstitial and vacancy loop densities there was found their  universal character \emph{versus} dose (irradiation time).  The universal regimes are described by  two growth exponents related to  fast  and slow growth  denoted by terminal doses when the loop number densities of two types of loops saturate.

It was shown that loop sizes increase from the grain boundary to the center of the grain. The same behaviour was observed for growth strain in $X$ and $Y$ directions of the Cartesian system; the growth strain values in $Z$ direction decrease from the edge of the grain to its center. This effect is observed only  at elevated accumulated doses. At doses up to $5$~dpa there is no inhomogeneity of growth strain distribution inside the grain and, therefore, the whole sample manifests radiation growth with the equivalent deformations in Cartesian coordinate system.

It was shown that the  mean size of dislocation loops and growth strains do not depend essentially on irradiation temperature at reactor conditions ($550{-}600$~K). Observable changes of loop radii (up to $3{-}5$~nm) and growth strains in $X$ and $Y$ directions were found only at temperatures up to $550$~K. It was found that the temperature growth does not change the dynamics of loop density.

The growth rate of dislocation loops at doses up to $23$~dpa (as the terminal dose for $\langle c\rangle$-type loops) and the loop size is defined by grain size and size of precipitates of secondary $\beta$-phase enriched  by Nb atoms from $80\%$ up to $90\%$. Dislocation loops grow faster in large grains; with an increase in precipitates size, the radius of  interstitial loops  takes elevated values; the size of $\langle c\rangle$-type loops slightly decreases.   

By studying the processes of loops growth at different content of niobium as an alloying element we have shown that observable (up to $2$~nm by TEM) $\langle a\rangle$-type vacancy loops which are of metastable character  emerge only at a low concentration ($<0.5\%$) of Nb and at low temperatures ($<573$~K). Their size is up to several nanometers, the dose range of their emergence is up to $3$~dpa depending of Nb content, grain and precipitate size, irradiation temperature. A decrease in Nb content increases the growth strains in $X$ and $Y$ directions and decreases their values in $Z$ direction only at elevated doses (above $5$~dpa). A temperature  decrease suppresses interstitial loops formation and growth, increasing the corresponding incubation dose.

The obtained results relate well to experimental observations of the defect structure formation in neutron irradiated zirconium based alloys (see reviews \cite{koenig1,Zr-Nb_dislo} and citations therein, \cite{NGB79,G88,H88,G93}, and discussions of dislocation structure evolution in OSIRIS rector \cite{PTG17}) and are in good correspondence with the previously obtained data from theoretical modelling \cite{GBS13,RRT,BGS15,PTG17}.  The proposed model generalizes the existing approaches of defect structure  dynamics and  can be used to study the evolution of defect structure in binary alloys under sustained irradiation.

\ukrainianpart

\title{Ріст дислокаційних петель та радіаційний ріст у сплавах Zr-Nb, опромінюваних нейтронами: моделювання в рамках теорії швидкостей реакцій}
\author{Л. Ву\refaddr{label1}, Д.О. Харченко\refaddr{label2}, В.О. Харченко\refaddr{label2}, О.Б. Лисенко\refaddr{label2}, В. Купрієнко\refaddr{label2}, І.О. Шуда\refaddr{label3}, Р. Пан\refaddr{label1} }
\addresses{
\addr{label1} Перший інститут, Інститут ядерної енергії Китаю, Перша секція, дор. Чангшундадао, 328, Шуангліу, 610213 Ченгду, Китай  
\addr{label2} Інститут прикладної фізики НАН України, вул. Петропавлівська, 58, 40000 Суми, Україна 
\addr{label3} Сумський державний університет, вул. Римського-Корсакова, 2, 40007 Суми, Україна
}

\makeukrtitle

\begin{abstract}
\tolerance=3000%

Представлена узагальнена модель росту дислокаційних петель в опромінених бінарних сплавах на основі Zr. Ураховано вплив температури, ефективність поглинання точкових дефектів петлями, що залежать від розміру петлі, локальність розташування петель в зернах і концентрації легуючого елемента. Розроблену модель використано для опису динаміки зростання радіусів дислокаційних петель  в цирконій-ніобієвих сплавах, опромінених нейтронами за реакторних умов. Встановлено динаміку  росту  як радіусів петель різного типу, так  деформацій кристалу  при різних розмірах зерен, розташуванні петель в зернах і концентрації ніобію. Показано, що залежно від розташування петель в зернах відбувається  нерівномірна деформація матеріалу всередині зерен. Показано, що  введення ніобію як легуючого елемента призводить до  зменшення  радіусу дислокаційних петель та сприяє зростанню локальних деформацій всередині зерен.
\keywords опромінення, цирконієві сплави, точкові дефекти, дислокаційні петлі, радіаційний ріст

\end{abstract}  
\end{document}